\documentclass[preprint,12pt,authoryear]{elsarticle}
\usepackage{amssymb}
\usepackage{amsmath}
\usepackage[hidelinks]{hyperref}
\usepackage{tabularx}
\usepackage{color}
\usepackage{booktabs}
\usepackage{ragged2e}
\usepackage{longtable}
\usepackage{xcolor}  
\usepackage{natbib}
\biboptions{authoryear,round}
\usepackage{placeins}

\journal{Computers in Human Behavior Reports}

\begin{document}

\begin{frontmatter}

\title{Human–AI Interactions and the Psychological Triad: A Multi-Domain Survey of Cognitive, Behavioral, and Emotional Impacts}

\author[inst1]{Omar Al-Refai\corref{cor1}}
\ead{omaralrefai@tamu.edu}

\author[inst2]{Yadira Colunga Reyes}
\ead{yadira.colunga\_reyes@students.tamuk.edu}

\author[inst2]{Celeste Riley}
\ead{celeste.riley@tamuk.edu}

\author[inst1]{Eman Hammad\corref{cor1}}
\ead{eman.hammad@tamu.edu}

\address[inst1]{Texas A\&M University, College Station, TX 77840, USA}
\address[inst2]{Texas A\&M University-Kingsville, RELLIS Campus, Bryan, TX 77807, USA}

\cortext[cor1]{Corresponding authors}

\begin{abstract}
As human-AI interactions increasingly reshape social and research landscapes, the risks of overreliance, cognitive offloading, and the nuanced degradation of human agency have become more apparent. This paper addresses these challenges by introducing a novel conceptual framework based on the psychological triad: cognition, behavior, and emotion. Through a comprehensive survey of recent research, we observe that while AI can substantially enhance outcomes such as memory, creativity, and engagement, it simultaneously introduces systemic risks, including diminished critical thinking, skill erosion, and increased anxiety. Emotional outcomes are shown to be similarly mixed, with AI systems offering support and stress reduction while raising critical concerns regarding dependency, inappropriate attachments, and ethical oversight. By synthesizing these findings into the TRIAD model, a recursive framework of appraisal, execution, and regulation, this paper moves beyond isolated metrics toward a systems-level understanding of how AI shapes cognitive depth, behavioral patterns, and emotional development over time. The proposed model underscores the need for responsible, context-aware AI integration and establishes a grounded evaluation framework capable of capturing the complex, human-centered trade-offs inherent in modern technological advancement.
\end{abstract}



\end{frontmatter}
\section{Introduction}
\label{sec:intro}
The rapid growth of artificial intelligence across the global socio-technical landscape has intensified the need for structured frameworks to examine its multi-dimensional impact on human agency. While AI systems are frequently evaluated through the narrow lenses of technical performance and operational efficiency, their true effects extend far beyond task optimization. These technologies fundamentally reshape the internal mental processes that govern how individuals interact, make decisions, and develop. This article addresses these complexities by introducing a systems-level analysis of recent research through the foundational lens of the psychological triad: cognition, behavior, and emotion. While recent surveys, such as~\citet{Williams2024PsychologyAI}, have utilized a similar basis within the specific contexts of management and marketing, this work expands that scope to provide a broader, interdisciplinary synthesis of human--AI interaction.

From an engineering and design perspective, the integration of AI across education, healthcare, and industrial systems influences not only individual psychology but also the fundamental architecture of human-centered applications. Engineering solutions traditionally aim to mitigate system failures and optimize user experience (UX) by leveraging human factors, system reliability, and performance metrics as validation criteria~\citep {Car10}. However, as AI systems become increasingly autonomous and opaque, these traditional models must evolve. This survey argues that understanding AI outcomes through the psychological triad is essential for developing a comprehensive foundation for ethical design and responsible technological policy. By analyzing the recursive trade-offs between enhancement and degradation, we provide a framework to guide the next generation of human-compatible AI.

The remainder of this paper is organized as follows: Section~\ref{sec:triad} establishes the theoretical foundation by introducing the psychological triad and the taxonomy used to categorize human--AI outcomes. Section~\ref{sec:cognitive} examines the cognitive impacts of AI, using Bloom's Taxonomy to contrast creative ideation with the risks of cognitive offloading. In Section~\ref{sec:behavioral}, we analyze behavioral shifts through the I-PACE framework, focusing on user agency and task execution. Section~\ref{sec:emotional} explores the emotional landscape, including the potential for therapeutic support versus risks like social withdrawal. Section~\ref{sec:analysis} synthesizes these dimensions by discussing the interconnected feedback loops within the psychological triad. Section~\ref{sec:triad_model} proposes the TRIAD model, which conceptualizes human--AI interaction as a recursive, triadic process unfolding across cognitive, emotional, and behavioral streams. Finally, Section~\ref{sec:discussion} provides a systems-level discussion on the future of human--AI interactions research, followed by concluding remarks in Section~\ref{sec:conclusion}.

\section{AI Impacts Along The Psychological Triad}
\label{sec:triad}
\begin{figure}
    \centering
    \includegraphics[width=0.76\linewidth]{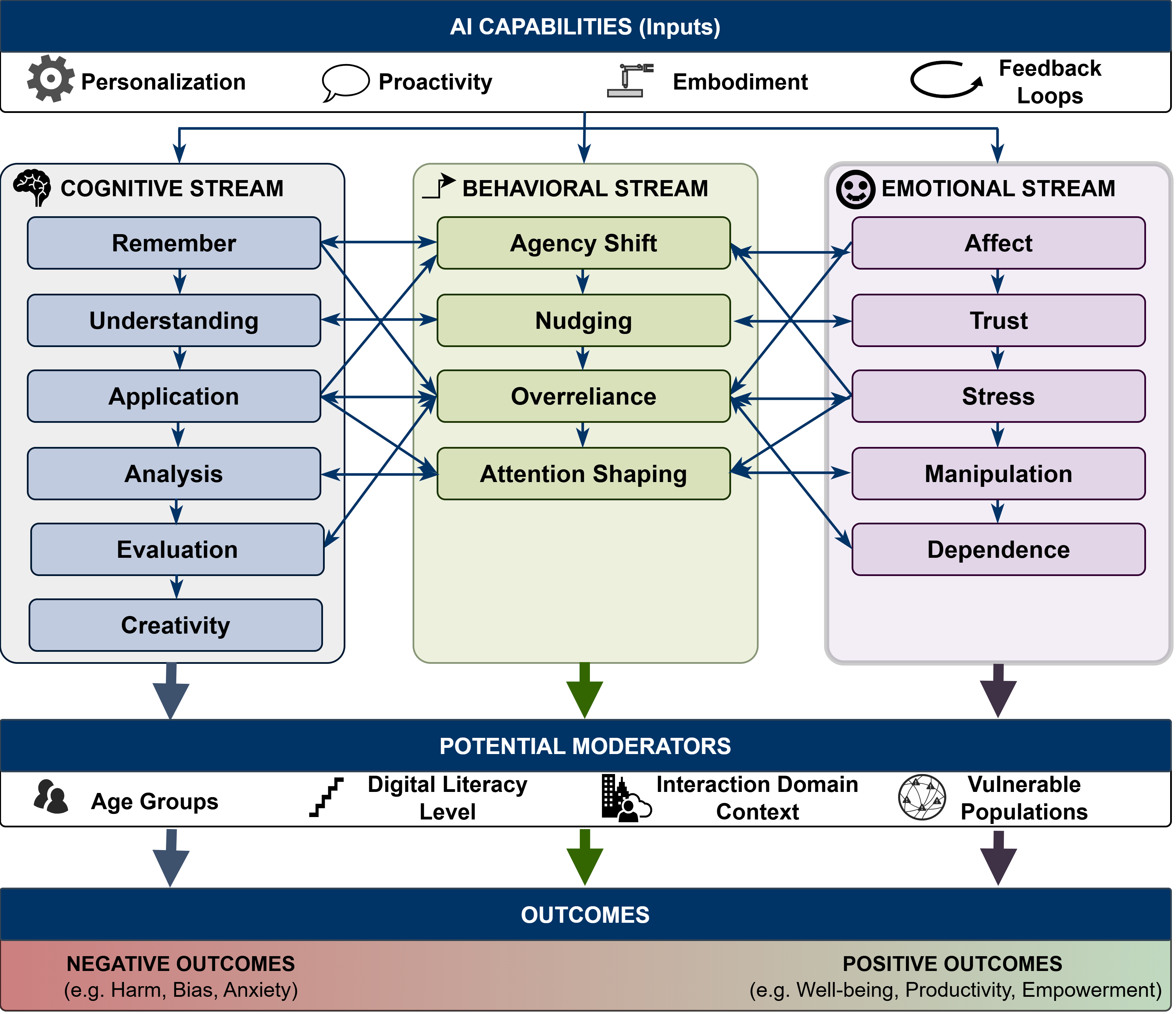}
    \caption{Illustration of the Human-AI Interaction concepts outlined in the survey.}
    \label{fig:papersummary}
\end{figure}

Several tripartite models exist in psychology; however, the trio of thoughts, emotions, and behaviors is widely used to succinctly describe the field of psychology and the study of personality. Originally, the connections among cognition, emotion, and behavior were explored to explain the components of individual attitudes. Here, we employ this tripartite structure to survey the human impacts of AI/ML, while mindful that human thoughts, feelings, and behaviors are not neatly siloed. More advanced, interactive models of human-AI outcomes will be highlighted to help identify knowledge gaps. The taxonomy and the structure of the remainder of the paper are outlined in Figure \ref{fig:papersummary}.

\section{Cognitive Impacts}
\label{sec:cognitive}
Cognition is defined in the psychological triad as the cognitive processes an individual uses to take in (encode), manipulate (process/modify), and make decisions based on information (use). As a result of recent advances in technology, these processes that were historically attributed to the human mind are now being externalized through artificial intelligence (AI) systems. The application of AI can be seen across a range of areas, including education and healthcare~\citep{Bab24}, due to its versatility as a tool~\citep{XuY21}. To sustain this level of adaptability, AI scientists have implemented a number of components in these systems that approximate certain aspects of human information processing, enabling models to learn from data and dynamically adjust their output over time~\citep{rawas2024ai}.  

\begin{figure}
    \centering
    \includegraphics[width=1.0\linewidth]{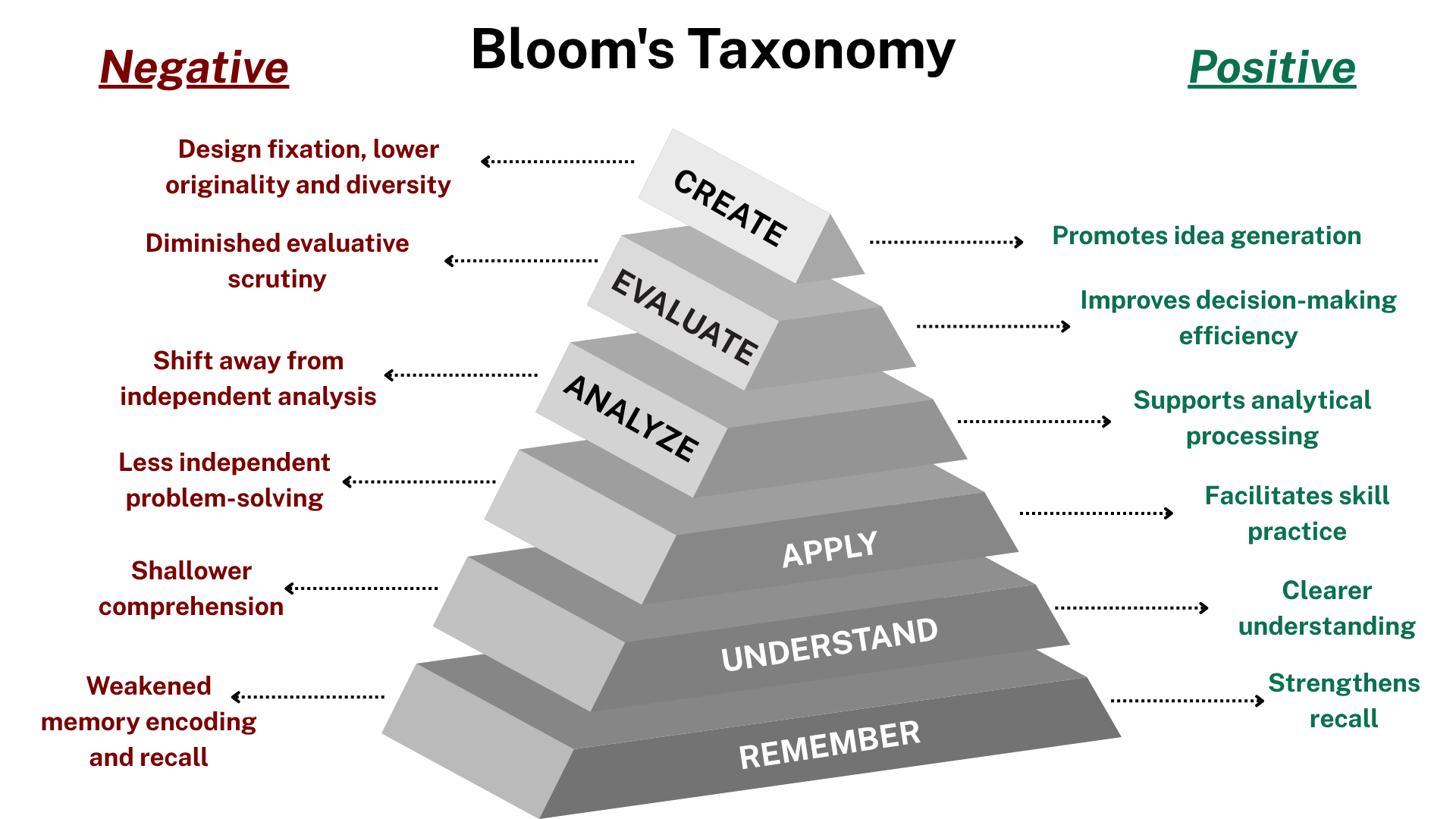}
    \caption{Cognitive effects of AI use across Bloom’s Taxonomy}
    \label{fig:blooms_taxonomy}
\end{figure}

Whereas AI systems are designed to adapt through iterative learning, human cognition is biologically elastic and responsive to environmental inputs. Consequently, users may experience changes in their cognitive processes due to long-term use of AI systems~\citep{Sin25}. However, the extent of these cognitive effects remains insufficiently understood. It is unclear how specific processes are being enhanced, redistributed, or attenuated through repeated AI engagement. This section draws on the previous work of \citet{Sin25} and applies Bloom’s Taxonomy (shown in Figure~\ref {fig:blooms_taxonomy}) to systematically examine how AI both enhances and constrains across lower and higher order levels of cognition. Although Bloom’s Taxonomy was originally developed as a pedagogical framework rather than a formal model of cognition, it provides a systematic hierarchical structure for differentiating how AI may affect the processes of remembering, understanding, application, analysis, evaluation, and creation~\citep{krathwohl2002revision}.  This framing facilitates the question of whether shifts in how individuals interact with lower-level cognitive tasks with AI assistance may cascade into effects on complex reasoning. Studies considered in this section are summarized in Table ~\ref{tab:cognitive}.

\subsection{Positive Cognitive Impacts}
\subsubsection{Memory and Recall (Remember)} 
Deep memory encoding is a critical component of learning new information. Research has consistently demonstrated that practicing active recall strongly enables this process \citep{rowland2014effect, Ibn25}. In educational contexts, learners often report awareness of the effectiveness of retrieval-based strategies but fail to implement them due to the burden of preparation and time constraints \citep{rea2022students, Ibn25}. Artificial intelligence may help address these barriers by providing accessible self-testing opportunities. As shown by \citet{Ibn25}, students who used AI-generated practice questions achieved significantly higher final exam scores than those who did not. Similarly, \citet{AnY25} found that exposure to LLM-generated practice was associated with higher knowledge retention (89\%) compared to no exposure (73\%). These differences in scores highlight the extent to which recall improved with repeated exposure facilitated by AI-supported retrieval practice. 

\subsubsection{Comprehension (Understand)} 
Grasping the meaning of an idea requires that information be cognitively accessible to the learner, but this is not always the case, as information is sometimes presented in complex forms. ChatGPT’s capacity to simplify material has been shown to improve students’ reading comprehension and inferential processing of text \citep{Cel24}. AI also offers flexibility through personalized learning, which can facilitate understanding. \citet{Hid24} examined the use of an AI-based personalized reading platform (ReadTheory) among high school students in Indonesia and found that those who practiced with it demonstrated higher reading comprehension than those in a control group. This personalization aligned with each learner’s needs by engaging at reading levels that were challenging without being overwhelming, thereby promoting more effective comprehension of the content.  

\subsubsection{Skill Application (Apply)} 
Developing strong skills and knowledge requires practical application of concepts. In a review on the use of artificial intelligence in STEM education, \citet{XuW22} explains that AI applications, such as educational robots and intelligent tutoring systems, allow students to apply knowledge through interactive learning experiences. Generative AI also offers clear gains within a flipped classroom model, which emphasizes student engagement rather than traditional teacher-centered instruction \citep{Kwa25}. In this model, students learn core content before class and complete assessments that identify gaps in their knowledge. This preparation allows class time to be used for applying learned concepts through active discussions. After class, AI can further reinforce learning by generating targeted instructional materials and providing immediate feedback. These functions illustrate how AI can help identify areas of weakness and continue development through application and reinforcement.  

\subsubsection{Analysis (Analyze)} 
Analyzing involves breaking information into components and critically examining patterns and relationships. Artificial intelligence can enhance analytical processing by rapidly identifying features across large datasets without the fatigue that humans inevitably experience. In medical imaging, \citet{Cha16} found that radiologists using an AI-aided detection system identified 13.8\% more breast cancers than those interpreting mammograms without AI assistance, indicating that algorithmic pattern recognition can help surface subtle abnormalities that might otherwise be overlooked. Evidence from fatigue and workload research further clarifies the cognitive basis of such perceptual errors. Prior work has shown that radiologists exhibit measurable deterioration in visual scanning behavior as mental fatigue accumulates, including reduced systematic coverage of images \citep{pershin2022artificial}. In this context, AI-assisted pattern recognition can help maintain consistent feature identification across large volumes of data, thereby alleviating sustained visual search demands and supporting the allocation of cognitive resources toward integrative analytical processing, such as connecting detected image features.

\subsubsection{Judgment and Evaluation (Evaluate)} 
Making decisions requires weighing evidence and judging among possible courses of action. This evaluative process becomes increasingly challenging as the number of available options and possible outcomes grows. In such situations, AI can function as an evaluative assistant by organizing alternatives and supporting comparison based on user-defined criteria. In particular, predictive AI demonstrates the capacity to provide structure for strategic evaluation by transforming large-scale data into anticipatory insights, enabling decision-makers to assess potential trajectories rather than relying solely on retrospective judgment \citep{Vud23}. For example, in a university Information and Communication Technology (ICT) department, \citet{Fun24} found that an AI-powered decision support system integrated operational data with predictive analytics, enabling IT staff to anticipate issues and prioritize interventions. Participants reported that the system was easy to use and that the information it provided helped them carry out operational decisions more efficiently. While these findings are based on self-reporting rather than direct cognitive measures, they suggest that AI support reduced the practical demands of evaluation overall.

\subsubsection{Idea Generation (Create)} 
Creating draws on existing knowledge and transforms it into original approaches. Emerging research suggests that generative AI can support novel directions during ideation. \citet{Sha24} investigated an AI-augmented group brainwriting method in an undergraduate Tangible Interaction Design course and found that integrating large language models into the ideation process increased the number of ideas generated and broadened the range of solutions considered. Similarly, \citet{Hab24} reported that participants using ChatGPT during ideation tasks demonstrated higher levels of divergent thinking, particularly in idea fluency and elaboration. This reflects a cognitive benefit associated with divergent thinking (the generation of multiple ideas) rather than convergent thinking, which involves narrowing toward a single solution.

\subsection{Negative Cognitive Impacts}

\subsubsection{Memory and Recall (Remember)}

While AI can strengthen recall from repeated exposure, evidence also highlights prospective cognitive costs associated with overreliance. \citet{Abb24} found that frequent adoption of ChatGPT in academic activities was associated with reported memory difficulties, including forgetting tasks and overall challenges in retaining learned information. Concerns about memory have been observed specifically when AI is integrated into writing tasks. In \citet{Kos25}, students wrote essays with or without large language model (LLM) assistance, followed by recall assessments. The results showed that individuals who used AI demonstrated poorer recall of what they had just written, whereas only a small proportion of those who wrote without AI demonstrated similar poor recall. These findings raise questions about how reliance on automated generation may influence memory processes over time. 

\subsubsection{Comprehension (Understand)}
Despite AI’s capacity to modify content and assist comprehension, the accuracy of generated outputs remains an important consideration. Language models tend to overgeneralize information drawn from source material, meaning they do not always provide precise representations of the original text \citep{Hak24, Pet}. This lack of accuracy can influence how learners interpret information, sometimes leading to incomplete or inaccurate understanding. Consistent with this concern, \citet{Etk25} found that individuals with high pre-existing comprehension skills demonstrated less accurate understanding when reading AI-generated summaries than when reading the original text. The authors explain that this may occur because AI-generated summaries often omit critical details required for high-quality comprehension. Thus, ensuring that AI use enhances rather than undermines information uptake requires verifying details from the original text.

\subsubsection{Skill Application (Apply)}
There is growing evidence that AI can function as a learning scaffold, though allowing it to take over cognitive effort in assignments raises the possibility of unintended consequences. Earlier research on automation has shown that when people allow technology to handle complex tasks, they may gradually lose accuracy and confidence in performing those tasks independently \citep{Vol16}. \citet{Bas25} found that while ChatGPT (specifically GPT-4) improved immediate mathematical problem solving, students performed worse on subsequent independent tasks than those in the control group. These observations highlight that sustained engagement is essential for maintaining conceptual knowledge and skills, and that AI may serve either as a mechanism for skill application or as a factor that gradually diminishes competence. It is also important to note that higher-order application depends on lower-order cognitive processes. Bloom’s Taxonomy positions recall and comprehension as foundational, with application and subsequent higher-level skills necessarily integrating these earlier forms of cognition \citep{adams2015bloom}.

\subsubsection{Analysis (Analyze)}
Alongside its role in supporting analytical reasoning, AI may reduce the extent to which individuals independently engage in analysis. \citet{gerlich2025ai} found that higher levels of AI tool usage were associated with lower performance on critical thinking assessments, signaling reduced direct cognitive involvement. Additionally, \citet{Lee25} surveyed knowledge workers whose roles depend on information processing to examine how they interacted with AI. The study found that when using generative AI, participants shifted cognitive effort away from active critical analysis toward managing and integrating AI output, particularly when users reported high confidence in the system. Such shifts suggest that users may bypass analytical processing and instead engage in an evaluation/supervisory cognition mode, situating analysis as a shared function between humans and the system.

\subsubsection{Judgment and Evaluation (Evaluate)}

Although AI can improve evaluative efficiency, it may prioritize information in ways that differ from human judgment. Personal values and contextual factors that matter to the individual may not be fully captured in AI-driven assessments. \citet{Der24} argue that AI chatbots differ fundamentally from earlier technological tools. Unlike passive systems, their conversational and interactive nature enables them to function as extensions of human cognition by actively participating in reasoning, interpretation, and judgment. This dynamic interaction may change how individuals judge information and how much they rely on AI outputs during decision-making. Moreover, the way that AI presents information can substantially alter how users evaluate the content it produces. Studies have found evidence that individuals often struggle to distinguish between passages written by humans and those generated by OpenAI's ChatGPT (GPT-4) model~\citep{Rip25}. These findings emphasize how AI-generated texts closely mirror human style, structure, and coherence, potentially leading users to defer to AI recommendations rather than rely on their own judgment and intuitive assessments~\citep{li2024developing}. 

\subsubsection{Idea Generation (Create)}

Idea generation is often viewed as a benefit of AI usage, though concerns persist that generative AI could promote uniformity and constrain originality. In an experimental study, \citet{Wad24} examined participants' use of the AI image generator Midjourney as a creative aid during a visual ideation task. They discovered that AI-assisted designs exhibited less variety and originality than those produced using Google Image Search or no external aid. The researchers further observed that there was frequent imitation of AI-generated visual elements, indicating design fixation; a phenomenon in which exposure to examples unconsciously limits the exploration of novel alternatives. Similarly, \citet{Dos24} found that when participants used AI to write stories, individual creativity increased; however, the resulting outputs became more similar across participants, reflecting a reduction in collective diversity within creative work.

\subsection{Mixed Cognitive Outcomes in Special Populations}
\subsubsection{Geriatrics}
AI may serve as a supportive cognitive tool for older adults when applied in structured conditions. \citet{Kan} examined an LLM-based AI telephone intervention designed to provide regular conversational engagement to the elderly with dementia. Following the intervention, participants demonstrated a significant improvement in memory function, whereas no significant changes were observed in attention or language. The authors further reported that memory-related benefits were more pronounced among female participants and varied by education, suggesting that individual characteristics may moderate the cognitive effects of AI-based interventions. Another study, \citet{AlR25} reported that a VR-AI immersive program incorporating virtual environments, memory-oriented tasks, and an AI conversational companion produced statistically significant improvements in cognitive screening performance, as measured by the Montreal Cognitive Assessment (MoCA). The authors note that these findings are preliminary and based on a pilot sample, but suggest potential positive cognitive outcomes from immersive AI-supported interventions. At a broader level, a systematic review and meta-analysis examining socially assistive AI robots reported moderate improvements in global cognitive screening outcomes among older adults \citep{Lee22}. Across studies, stronger effects were observed for anthropomorphic designs compared to non-humanoid systems, with humanoid robots, such as Sil-bot included, among the more effective interventions. These effects may signal a positive contribution of human-like AI features and their potential to enhance cognitive engagement among older adults. 

\subsubsection{Learning Disabilities}
Generative AI tools have expanded access to educational assistance for students with disabilities, opening opportunities that may not have been previously available. A survey of higher-education students with disabilities, including ADHD, dyslexia, dyspraxia, and autism, revealed that ChatGPT was a commonly used application during periods of difficulty sustaining attention or organizing their thoughts~\citep{Zha25}. Within a broader disability sample, \citet{Pie25} found that students with visual and hearing impairments frequently relied on AI for reading comprehension and writing support. They reported using ChatGPT to summarize texts, structure assignments, and manage language-related difficulties. At the same time, the study highlights emerging concerns regarding overreliance and skill erosion. For instance, one dyslexic student questioned the extent to which spelling skills should still be learned when an AI system can automatically correct errors~\citep{Pie25}. Overall, generative AI appears to reduce barriers for students with disabilities, but its use may also shift learning away from sustained cognitive skill practice toward greater reliance on automated assistance. 

\subsection{Mitigations}
To maximize the benefits of AI while mitigating cognitive risks, effective human-AI collaboration requires active user engagement rather than passive reliance. \citet{lim2025debiasme} suggests that mitigation depends not only on system design but also on users deliberately adopting metacognitive strategies during AI use. Specifically, users must (a) pause to reflect on how prompts are framed and whether they embed prior assumptions, (b) critically evaluate AI-generated outputs rather than accepting them as authoritative, and (c) engage with scaffolding that adapts to task demands and stages of interaction rather than treating AI support as uniform across contexts. By engaging with AI as a cognitive partner rather than an automated solution, users retain agency over reasoning processes and reduce unreflective dependence. While these metacognitive strategies are theoretically grounded and well aligned with existing cognitive bias research, their effectiveness in reducing cognitive offloading and sustaining critical engagement requires further empirical evaluation across diverse users, tasks, and contexts.

\subsection{Discussion}
As AI becomes regularly used in educational, professional, and assistive settings, understanding the influence on cognitive functioning becomes increasingly important. Evidence reviewed across cognitive domains indicates that AI can meaningfully support thinking processes by reducing access barriers, organizing complex information, and augmenting performance across multiple levels of cognitive demand. However, these advantages are not universal and appear highly contingent on context, task structure, and patterns of use. While AI-supported tools show promise for enhancing recall, comprehension, and idea generation, converging findings also point to potential tradeoffs related to reduced independent engagement and the redistribution of mental effort. Cognitive work appears to shift from actively performing reasoning to supervising, verifying, or integrating AI-generated output. Such shifts raise critical questions about how sustained reliance on AI may affect long-term skill development, particularly in fields where expertise depends on repeated practice, deep domain knowledge, and independent judgment. If foundational cognitive skills are increasingly delegated to automated systems without continued human practice, it becomes unclear how future experts will develop, maintain, and refine the competencies required for high-level professional performance. These observations underscore the importance of determining not only whether AI can serve as a resource for cognition, but how it can be integrated without displacing essential skill-building processes.  

Future research should move beyond short-term performance gains to examine how sustained AI use changes cognitive development over time. In particular, studies are needed to clarify whether delegating lower-level cognitive processes to AI systems frees cognitive resources for reallocation to higher-order reasoning or whether reduced active engagement in foundational cognitive work undermines the encoding, consolidation, and transfer of knowledge, thereby constraining higher-level thinking. Greater attention is also warranted to individual differences, such as age, educational background, and learning needs, that may have different cognitive outcomes. Ultimately, responsible integration of AI into cognitive domains depends not only on efficiency or accuracy, but on preserving the conditions under which human expertise is cultivated, sustained, and advanced alongside intelligent technologies. 


\begin{table*}[htbp]

\caption{Summary of AI Impact Studies: Cognitive Focus}
\label{tab:cognitive}
    \centering
    \tiny
    \renewcommand{\arraystretch}{0.95}
    \setlength{\tabcolsep}{3pt}
    \begin{tabularx}{0.90\textwidth}{
    >{\RaggedRight\arraybackslash}p{1.6cm} >{\RaggedRight\arraybackslash}p{2.7cm} >{\RaggedRight\arraybackslash}p{1.5cm} >{\RaggedRight\arraybackslash}p{1.2cm} >{\RaggedRight\arraybackslash}p{4.0cm}}
        \toprule
        \textbf{Impact Focus} & \textbf{Study Topic / Focus} & \textbf{Study Type} & \textbf{Sample Size (N)} & \textbf{Key Metrics / Questions Used} \\ 
        \midrule
        Knowledge Acquisition & ChatTutor vs. ChatGPT for Learning \citep{Mak25} & Between-subjects experiments & N=175 (Uni); N=234 (HS) & 9-item conceptual knowledge test; Likert scales (1-5) for self-efficacy, trust, and enjoyment. \\ 
        Critical Thinking Skills & AI and Critical Thinking Skills \citep{gerlich2025ai} & Mixed-method (Survey + Int.) & N=666 (Sur.); N=50 (Int.) & Halpern Critical Thinking Assessment (HCTA); Terenzini’s measures of development. \\ 
        Memory & Alzheimer’s VR-AI Intervention \citep{AlR25} & Comparative pilot study & N=23 (Int.); N=10 (Ctrl) & Montreal Cognitive Assessment (MoCA); task completion rate; caregiver feedback. \\ 
        Knowledge Retention & ChatGPT-Assisted Retrieval Practice \citep{Ibn25} & Retrospective cohort study & N=128 & Final exam scores (60 MCQs); Feedback questionnaire on AI feedback quality. \\ 
        Knowledge Retention & LLM-Generated Retrieval Practice \citep{AnY25} & Quasi-experiment & N=58 & Quiz accuracy (percentage correct) comparing weeks with/without LLM questions. \\ 
        Skill Degradation & Cognitive Skill Degradation (Aviation) \citep{Vol16} & Between-subject experiment & N=59 (Init); N=46 (Final) & NASA-TLX (subjective workload); objective error rates in flight planning. \\ 
        Reading Comprehension & Reading Comprehension with GPT \citep{Etk25} & Randomized cross-over study & N=195 & ACT quiz accuracy; timed measures for passage reading; perceived effectiveness. \\ 
        Memory & AI Care Calls for Dementia \citep{Kan} & Longitudinal study (31 wks) & N=80 & Cognitive Impairment Screening Test (CIST); Geriatric Depression Scale (GDS). \\ 
        Risk Detection & Suicidal Ideation Detection \citep{chan2022two} & Technical validation & N=318,797 & Accuracy, precision, recall, and F1-score based on the Reddit Suicide Watch dataset. \\ 
        Decision Making & AI Decision Support in ICT \citep{Fun24} & Case study / Questionnaire & N=28 & Constructs for Decision Support, Efficiency, Integration, User satisfaction, and Scalability. \\ 
        Self-Efficacy & AI Writing Software Adoption \citep{papakonstantinidis2024embrace} & Quantitative survey & N=219 & Technology Acceptance Model (TAM) and New Computer Game Attitude Scale (NCGAS). \\ 
        Learning Quality & Emotionally Enriched AI Feedback \citep{alsaiari2024emotionally} & Randomized experiment & N=381 & AEQ-S (41-item emotion survey); 5-point helpfulness ratings; interaction time. \\ 
        Correctness Evaluation & AI vs. Human Reading Passages \citep{Rip25} & Mixed-methods & N=89 & Likert scales (1-5) for readability, correctness, coherence, engagement, and adequacy. \\ 
        Learning Performance & GPT Base vs. Guardrailed GPT Tutor \citep{Bas25} & Randomized controlled trial & N=839 & Grades on assisted practice vs. unassisted exams; average message count per problem. \\ 
        Creative Novelty & AI and Individual/Collective Creativity \citep{Dos24} & Preregistered experiment & N=293 (W.); N=600 (E.) & 9-point scale indices for novelty and usefulness; emotional state; ownership/profit split. \\ 
        Cognitive Bias & AI Feedback Loops and Human Bias \citep{glickman2025human} & Behavioral experiments & N=1,401 & Decision change percentages; confidence ratings; weight assignment in joint responses. \\ 
        Analytical Reasoning & AI and Human Loss in Decision Making \citep{ahmad2023impact} & Quantitative survey & N=285 & Likert scales for human laziness, loss in decision-making capabilities, and safety/privacy. \\ 
        Reading Comprehension & AI Simplification in EFL Reading \citep{Cel24} & Within-subjects intervention & N=105 & Reading comprehension test (RCT); Foreign Language Reading Anxiety Scale (FLRAS). \\ 
        Memory & Neural Costs of Essay Writing with AI \citep{Kos25} & Longitudinal crossover & N=54 (Init) & EEG connectivity (dDTF); NLP metrics; post-task interview quoting ability. \\ 
        Theory of Mind & Anthropomorphism / Perspective Taking \citep{zhao2022spontaneous} & Factorial design & N=1,431 & Level-2 VPT rates; 7-point Likert scale for mental capacity; 100-point humanlikeness slider. \\ 
        Cognitive State & AI and Robotic Social Empathy \citep{Lee22} & Systematic review & N=~575 & Mini Mental State Examination (MMSE); APADEM-NH (Apathy scale); AER Instrument. \\ 
        \bottomrule
        \end{tabularx}
\end{table*}


\section{Behavioral Impacts}
\label{sec:behavioral}

Taken together, the behavioral impacts of AI can be broadly understood through the lens of the I-PACE model, which provides a structured way of analyzing how person-level predispositions, affective responses, cognitive processing, and executive control interact with AI systems. Positive impacts arise when AI scaffolds cognition and strengthens self-regulation. Negative impacts emerge when AI displaces human agency, leading to skill degradation and overreliance. Neutral or mixed outcomes highlight the delicate balance between personalization and control. Ultimately, the behavioral trajectory depends not only on the technology but also on how it interacts with individual predispositions and emotional-cognitive processes, underscoring the need for responsible design that balances human agency with AI autonomy. 

\begin{figure}
    \centering
    \includegraphics[width=1.0\linewidth]{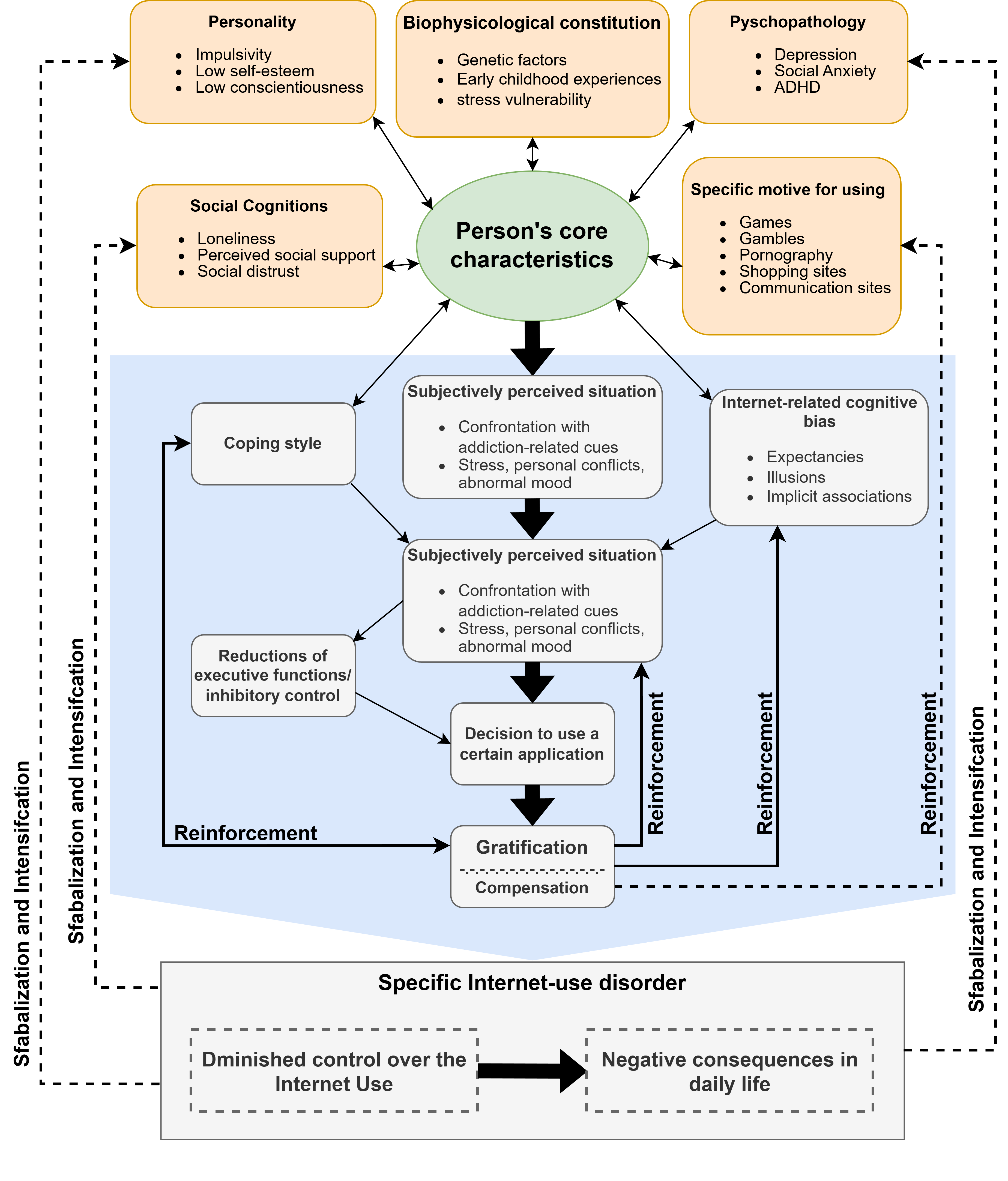}
    \caption{The I-PACE model, illustrating the mechanisms for the development and maintenance of Internet-use disorders proposed by~\citet{Bra16}.}
    \label{fig:I_PACE_MODEL}
\end{figure}

The Interaction of Person–Affect–Cognition–Execution (I-PACE) framework, shown in Figure \ref{fig:I_PACE_MODEL}, provides a unifying explanation of how person predispositions, affective reactions, cognitive processes, and executive control interact to drive technology use behaviors~\citep{Bra16}. Originally developed to explain the onset and maintenance of Internet-use disorder, the model emphasizes that behavioral consequences stem from the dynamic interplay between individual factors (e.g., neurobiological or psychological vulnerabilities), situational cues inducing affective and cognitive evaluations, and the individuals' degree of inhibitory and executive control they can muster in these situations. While the model was first conceived within problematic uses such as gaming or gambling, explanatory potential is more transferable to human–AI interactions. In this subsection, the I-PACE model is connected to the behavioral impact of AI, highlighting positive, negative, and mixed trends, thus placing current human–AI behaviors within an established psychological and neuroscientific framework. A summary of the behavioral studies discussed in this section is provided in Table \ref{tab:behavioral}. 

\subsection{Positive Behavioral Outcomes}
\subsubsection{Behavior Change via Personalization} 
The improvement in adaptive AI systems, such as DreamBox, enables them to customize instructions and content for each individual student based on their pace and skill level~\citep{DreamBox2024}. According to \citet{Dey23}, this leads to improved engagement with the material and significant correlations between AI engagement and reported behavioral modifications, suggesting the potential of AI to foster positive behavioral changes. Further, recent studies show that AI-powered tools may assist individuals in chronic disease management (e.g., Lark's real-time dietary and exercise feedback \citep{Lee20}, mental health (e.g., Woebot chatbot delivers cognitive behavioral therapy \citep{Fit17}, and addiction recovery (e.g., LifeDojo's customized relapse prevention strategies), which is effectively driving sustainable positive behavior change through tailored interventions \citep{Tim25}.

\subsubsection{Behavior Change via Nudging}
AI-driven nudging utilizes "just-in-time" adaptive interventions (JITAIs) to prompt users toward positive actions. AI-based systems such as Replika demonstrate how conversational AI can function as a persistent behavioral nudge~\citep{Replika2026}. Users who interact with Replika report increased self-reflection and emotional regulation, demonstrating how AI can reinforce adaptive coping strategies in daily life \citep{TaV20}. Similarly, AI-driven platforms like "Carrot Rewards" have demonstrated that micro-incentives tied to smaller daily actions can collectively foster healthier long-term habits \citep{Bro20}. Both examples map onto the I-PACE framework’s emphasis on how environmental cues interact with user cognition and affect. In early stages, these small nudges shape attention and reward expectations, while in later stages, they can help reinforce larger habits that become deep-rooted, but in a positive rather than compulsive manner.

\subsubsection{Task Performance} 
In clinical and professional settings, AI assists in task execution by managing routine behaviors. Studies on VR-AI applications for geriatric care show that "intelligent conversational partners" can drive increased engagement in memory-stimulating behaviors that participants might otherwise avoid due to frustration \citep{AlR25}. 

In the I-PACE framework, the Execution (E) component refers to the behavioral decision-making and implementation phase. When interacting with AI, this execution is no longer purely human; it becomes a shared performance outcome. In Cognitive-Behavioral Scaffolding, the AI serves as a "behavioral scaffold," reducing the cognitive load required for execution. For instance, the \citet{AuraHealth2025} and \citet{TheMindCompany2026Balance} apps guide users through mental wellness routines, essentially automating the "planning" phase of behavior so the user can focus solely on the "execution" of wellness tasks. Additionally, task performance is heavily influenced by the Cognition (C) component—specifically, user expectations. Research suggests a "placebo effect" where users perform better simply because they believe the AI is assisting them, regardless of the AI's actual contribution. Further, as tasks become more complex, AI can trigger either problem-focused coping (leading to innovative execution) or emotion-focused coping (leading to avoidance or passive execution), depending on whether the user perceives the AI as an "opportunity" or a "threat".

\subsection{Negative Behavioral Outcomes}
\subsubsection{Overreliance and Skill Degradation} 
The behavioral impacts of AI use can be observed in both the operational and human-centered design of systems. One major impact is Automation Overdependence. For instance, engineers working in environments with AI-driven machinery may develop reduced proactive behavior, relying on AI for routine decisions (routine erosion) or problem-solving (loss of agency). A case in point is the popular Boeing 737 Max incidents, where pilots' reliance on automated systems contributed to failure to act appropriately in the face of system malfunctions, leading to tragic outcomes~\citep{Spi21}. This highlights the importance of designing AI systems that foster human engagement and accountability, particularly in high-stakes environments.  

From an I-PACE perspective, overreliance on automation arises when cognitive biases and reduced inhibitory (restrictive) control are reinforced over time. If the system consistently suppresses the need for human judgment, the execution (E) stage of behavior eventually becomes automated and habitual, reducing the individual’s ability to detect and intervene during anomalies. Over time, this aligns with the later stage of the model, where diminishing executive functioning creates vulnerability to dysfunctional, and sometimes even catastrophic, outcomes~\citep{Bra16}.

\subsubsection{Attention Shaping via Social Media Feed}
Engagement-based ranking is a way platforms sort and recommend posts so that items likely to draw the most interaction come first. The system is trained to favor measures such as clicks, time spent, reactions, and resharing, even when those priorities conflict with goals like accuracy, viewpoint diversity, or user well-being. These ranking systems learn over time which items are most likely to draw attention and trigger emotional reactions, and this learning shapes what users keep seeing and interacting with again and again. In the I-PACE framework, these ranking functions work as steady external cues that shape behavioral reactions before deliberate control processing begins. Engagement-optimized ranking can shift attention (behavior) and emotion at scale. A large Facebook experiment showed that altering exposure to positive or negative posts changed the emotional tone of what people wrote \citep{Kra14}. This is particularly consistent with I-PACE, where certain cue-exposure shapes affect cognition, which then guides behavior. 

Controlled field experiments conducted during the 2020 US election demonstrate that feed choices and reshare pathways influence what people see and how information flows \citep{Dur23}. Effects on attitudes were mixed, yet exposure patterns changed in measurable ways. These dynamics can seed persistent habits and attentional biases within the I-PACE cycle. Further, large-scale audits of YouTube reveal a shift from milder channels to more extreme communities, showing that some recommendation paths connect these spaces~\citep{Rib20}. Such pathways can heighten cue reactivity and reduce stimulus-specific inhibitory control, leading to maladaptive execution over time. 



These dynamics are further compounded by interface dark patterns that steer users toward intrusive defaults and limit informed consent~\citep{Gray2018}. By employing "dark nudges" that reduce perceived control and erode user autonomy, platforms facilitate the I-PACE model's identification of degraded executive control. Originally developed to explain behavioral addictions~\citep{Bra16}, the I-PACE framework is highly relevant to AI-driven "attention shaping." Specifically, AI platforms leverage these patterns to trigger Affective (A) responses—such as the "fear of missing out" (FOMO) or a craving for social validation—to drive compulsive Execution (E), such as infinite scrolling. Over time, the repeated association between the AI stimulus and the rewarding behavioral execution becomes a conditioned habit. This process effectively bypasses the inhibitory control typically present in the Execution phase of the I-PACE cycle, transforming deliberate interaction into a maladaptive, automated response~\citep{DING2025149868}.

\subsection{Mixed Behavioral Outcomes}
\subsubsection{Shift in Human-AI Interaction}
\citet{Sun20} describes tensions between machine agency (AI autonomy) and human agency (user control) as central to Human-AI Interaction (HAII). Modern research emphasizes human interaction with technology itself, especially as AI becomes more agentic (e.g., smart assistants, algorithms). Traditionally, the media were controlled by human sources (journalists). Now, AI allows users to customize, curate, and create content, thereby altering the classical behavioral feedback loop. 

\subsubsection{Personalization vs. Customization} 
Personalization occurs when AI customizes content without direct user input (e.g., Netflix recommendations). When users actively change settings (by selecting news topics), customization takes place. In the I-PACE model, the distinction directly affects the interaction between affect and cognition. Bypassing conscious decision-making, personalization may automatically shape reward expectancies and attention biases. Customization, on the other hand, keeps users engaged during the cognitive stage and strengthens their sense of control and agency~\citep{Kni15}.

\subsubsection{Proactive vs. Reactive AI} 
Reactive personalization (opt-in) is preferred by users over proactive AI choices. AI systems can be classified as either proactive or reactive based on how they make decisions. Proactive AI acts on its own initiative, whereas reactive AI acts only when specifically instructed. User satisfaction, control, and trust are all greatly impacted by this discrepancy~\citep{Die25}. 

Based on the I-PACE model, reactive AI enhances intentional cognition and promotes healthier behavior execution, since users keep control over when they make decisions, choose actions that better match their stated goals after some reflection, and are less likely to react automatically or impulsively to cues initiated by the system. On the other hand, proactive AI may increase affective resistance and reduce inhibitory control by introducing unexpected cues, leading to fragmented attention, reduced task persistence, and lower compliance with system guidance. 

Proactive versus reactive engagement patterns are further clarified by the agency account proposed by \citet{Sun20}. While \citet{Mal23} provides evidence for AI’s involvement in both proactive and reactive learning processes, these findings are best understood through the lens of user control. Under the I-PACE framework, proactive AI often introduces unexpected cues that tax the user’s self-regulatory resources, potentially triggering impulsive affective responses. Conversely, reactive flows support intentional cognitive processes by maintaining the user’s sense of agency. By aligning the system’s behavior with the user’s explicit goals, reactive interfaces minimize the "tensions" between machine and human agency identified by \citet{Sun20}, ultimately fostering more disciplined and persistent engagement. 

\subsection{Considerations for Special Populations}
The behavioral effects of interacting with AI differ across groups of people, rather than being the same for everyone. Groups defined by age, neurodiversity, socioeconomic background, or job-related exposure tend to show different patterns of behavior when AI is introduced. These differences often reflect uneven cognitive resources, different motivational patterns, and the opportunities a person’s environment provides, all of which the I-PACE framework treats as key factors shaping outcomes. In these groups, AI can support helpful behavior patterns or reinforce harmful ones, based on the setting and how the system is designed.

\subsubsection{Older Adults and Behavioral Adaptation}
In older adults, AI-based systems such as voice assistants or care robots can reinforce independence and daily functioning by serving as behavioral scaffolds; reminders for medication adherence, physical activity, or social engagement \citep{an2025autonomyolderadultagentinteraction}. However, overreliance may lead to "delegation drift," in which habitual behaviors (e.g., navigation, scheduling) become externally regulated. Within the I-PACE model, this reflects a shift in the \textit{Execution} (E) phase: AI-mediated execution replaces self-initiated action. Mitigation strategies include “graduated autonomy design,” where systems periodically reintroduce manual tasks or encourage user-initiated actions to maintain behavioral agency.

\subsubsection{Children and Adolescents: Formation of Behavioral Habits}
For younger populations, AI systems (e.g., recommendation engines, adaptive learning tools, or conversational agents) interact with developing inhibitory control systems. In this context, emotional cues and quick feedback loops shape the formation of habits early on. Spending too much time in algorithmically curated spaces may encourage externally cued behavior, where users act mainly in response to platform prompts rather than their own goals. From the I-PACE perspective, this can be understood as an early increase in emotional and cognitive reactivity, alongside weaker executive control. One mitigation approach is to add transparent feedback loops to AI systems. In practice, this means showing users their usage patterns, adding prompts that slow immediate reward-seeking, and supporting simple self-monitoring habits~\citep{trustworthyaieductionchildren}.

\subsubsection{Children with Developmental Disabilities: AI-Augmented Behavioral Regulation}
Recent work in AI-augmented applied behavior analysis (AI-ABA) has opened new pathways to understand and support behavioral regulation in children with autism spectrum disorder (ASD) and other developmental disabilities~\citep{Ghafghazi_2021}. These groups often have ongoing difficulties with self-regulation, communication, and everyday adaptive skills, so they may need long-term support that is planned and tracked by applied behavior analysts (ABAs). Traditional ABA often relies on hand-recorded data and clinician judgment; while it can work well in practice, it requires substantial staff time and can lead to inconsistent records and differing interpretations among practitioners. AI-augmented systems can change how we study behavior by tracking it closely and providing feedback that updates reinforcement patterns in real time as new data comes in.

The AI-ABA framework combines data from multiple sources, such as speech, facial expressions, and movement, to support real-time, individualized learning environments and to match interventions to each person's needs. As illustrated in the system architecture presented by~\citet{Ghafghazi_2021}, multimodal input from wearables, cameras, and virtual environments feeds into an AI-driven “treatment personalizer” that generates individualized behavioral plans. These systems can detect repetitive behaviors, attention lapses, or agitation through body tracking, EEG, and physiological sensing, then deliver reinforcement cues or task modifications that guide users toward desired behavioral outcomes~\citep{Ghafghazi_2021}. Within the I-PACE model, this constitutes an enhanced feedback loop linking the \textit{Affective} and \textit{Execution} components, in which AI serves as a real-time moderator of cue exposure and behavioral response.

From a behavioral perspective, AI-mediated systems can support self-regulation by guiding users’ attention and participation, using reinforcement that adjusts to their responses. For example, reinforcement learning approaches built into virtual or augmented reality environments may help reduce problem behaviors, such as aggression, avoidance, or repeated off-task actions (e.g., repetitive motions, agitation) toward structured play or communicative acts~\citep{Ghafghazi_2021}. The same level of precise feedback that can make behavior more consistent may also lead people to rely too much on outside prompts unless that feedback is gradually reduced through planned fading protocols. 

One way to address this is to use adaptive fading, where the AI gradually scales back its intervention as the user’s self-started behaviors become steady, so the person keeps control while the behavior changes remain in place. Transparent, explainable AI (XAI) components are important for clinical accountability because they enable practitioners to see \textit{why} specific behavioral cues or reinforcers were chosen and applied~\citep{XAIACM, XAIMRI25}.

Overall, AI-augmented behavioral analytics (AI-ABA) can be useful for studying special populations, but it also brings practical and methodological challenges. When designed well, these systems provide structured support that helps people stay engaged, monitor their own progress, and build new skills. Without adaptive regulation, they may encourage people to rely too heavily on algorithmic feedback.

Recent evidence from a large-scale survey of over 1,000 Dutch citizens demonstrates that disparities in AI-related knowledge, skills, and attitudes have produced a new form of inequality—the \textit{AI divide}, which mirrors but extends beyond the traditional digital divide~\citep{AI_divide}. \citet{AI_divide} identified five user clusters based on AI competence: \textit{expert advocates}, \textit{expert skeptics}, \textit{average users}, and two particularly vulnerable groups, \textit{unskilled skeptics} and \textit{neutral unskilled}. The final two groups show the weakest understanding of AI and the least ability to work with it, even though they still spend a lot of time each day on news and entertainment platforms shaped by algorithms. They tend to be older, have lower levels of formal education, and have fewer skills for managing digital privacy, leaving them more open to behavioral manipulation and automation bias~\citep{AI_divide}.

In the I-PACE framework, low-literacy users who are frequently exposed sit at a key point where limited cognitive understanding and weak control over affect-based trust interact with strong situational prompts and automated reinforcement. Since these users struggle to understand what the algorithm is trying to do or adjust personalization settings, their behavior (E) is mainly shaped by AI-selected feedback loops rather than by their own self-directed decisions. Over time, this pattern may encourage people to become more passive in their behavior, to take part in routine ways without much thought, and to be more open to nudges that play on expected rewards or quick emotional reactions. Low AI literacy can limit a person’s ability to make informed choices online and make them more likely to react to prompts and cues, which may increase the spread of misinformation, deepen filter-bubble exposure, and encourage compulsive media use~\citep{AI_divide}. 

Empirical findings from recent studies suggest that AI competence and privacy protection skills tend to support each other. People with limited ability to manage their privacy are often grouped as low-competence users, meaning they often lack practical habits to protect themselves against manipulative recommendation systems or unclear forms of personalization \citep{AI_divide}. This fits the I-PACE model, which views executive control as a protective factor that can interrupt maladaptive cue–response loops. From a behavioral perspective, when people cannot track or adjust the feedback algorithms provide, it can weaken self-control, making actions more automatic and more influenced by external cues.

\subsection{Mitigations}
To reduce these behavioral differences, two related interventions, designed to work together, are recommended. Explainable AI (XAI) and transparent interfaces may reduce users’ dependence on recommendation systems by clarifying how the system arrives at its outputs, so end users can see the cause-and-effect links behind suggestions and regain the awareness needed for self-regulation. Second, AI literacy and privacy skills should be part of lifelong digital education so that people with less formal schooling or limited technical experience can understand, question, adjust, and resist automated prompts. These measures address the I-PACE Cognition–Execution gap by building in more reflective thinking before a person acts.

Overall, people who are frequently exposed to AI tools but have limited AI literacy may be especially vulnerable to misunderstandings, overtrust, or misinterpretation in AI-mediated settings. Their interactions with AI are not inherently dysfunctional; the issue is that guidance becomes one-sided, leading behavior to shift from deliberate, planned actions to mostly reactive responses. Within the I-PACE framework, these outcomes seem to reflect design features of AI systems, such as unequal exposure to cues, limited feedback control, and reduced involvement of executive processes, rather than problems rooted in the user. 

In practice, AI-based behavior-change tools can yield larger gains for some special populations, but also increase the risk of harm or unintended side effects. Without planned teaching and design steps, these users may end up being guided by AI instead of guiding it, which can reinforce unequal patterns of behavior across large groups. To reduce these effects, we need behaviorally adaptive AI that provides temporary external structure and then gradually returns control to the user, supporting self-regulation and deliberate action.

\begin{table}[htbp]
    \centering
    \tiny
    \caption{Summary of AI Impact Studies: Behavioral Focus}
    \label{tab:behavioral}
    \begin{tabularx}{\textwidth}{>{\RaggedRight\arraybackslash}p{2cm} >{\RaggedRight\arraybackslash}p{2.5cm} >{\RaggedRight\arraybackslash}p{1.8cm} >{\RaggedRight\arraybackslash}p{1.2cm} >{\RaggedRight\arraybackslash}p{4.5cm}}
        \toprule
        \textbf{Impact Focus} & \textbf{Study Topic / Focus} & \textbf{Study Type} & \textbf{Sample Size (N)} & \textbf{Key Metrics / Questions Used} \\ 
        \midrule
        Dependence & AI Dependence and Mental Health \citep{Hua24} & Longitudinal cohort study & N=3,843 & 5-item AI dependence scale; PHQ-9; GAD-7. \\ 
        Engagement & Alzheimer’s VR-AI Intervention \citep{AlR25} & Comparative pilot study & N=23 (Int.); N=10 (Ctrl) & Task completion rate; caregiver feedback; MoCA scores. \\ 
        Adherence & AI Virtual Coach for Perfectionism \citep{Joh24} & Randomised feasibility trial & N=85 & Adherence rates; DASS-21 (stress/anxiety); EDEQ-7. \\ 
        Communication Speed & "Smart Replies" in Communication \citep{Hoh23} & Randomized experiments & N=1,036 & Messages per minute; Perceived Cooperative Communication scale; VADER sentiment. \\ 
        Productivity & Workplace Psychological Safety \citep{american20242024} & Online survey (Harris Poll) & N=2,027 & Likert scales (1-4) on job satisfaction and performance self-ratings. \\ 
        Efficiency & AI Decision Support in ICT \citep{Fun24} & Case study / Questionnaire & N=28 & Constructs for Decision Support, Efficiency, Integration, and Scalability. \\ 
        Ethical Judgement & How Humans Judge Machines \citep{hidalgo2021humans} & Hypothetical scenarios & N=~6,000 & 7-point Likert scale measuring moral foundations: Harm, Fairness, Loyalty, etc. \\ 
        Diagnostic Accuracy & AI for Breast Cancer (AI-STREAM) \citep{Cha16} & Prospective cohort study & N=24,543 & Cancer Detection Rates (CDRs), Recall Rates (RRs), PPV1, and malignancy assessment. \\ 
        Technology Adoption & AI Writing Software Adoption \citep{papakonstantinidis2024embrace} & Quantitative survey & N=219 & Technology Acceptance Model (TAM) (intrinsic motivation, trust). \\ 
        Engagement & Emotionally Enriched AI Feedback \citep{alsaiari2024emotionally} & Randomized experiment & N=381 & Behavioral interaction time; resource quality scores; AEQ-S survey. \\ 
        Usage Behavior & GPT-4o and User Psychosocial Outcomes \citep{phang2025investigating} & Longitudinal RCT (28 days) & N=981 & Problematic Use Scale (PCUS); Lubben Social Network Scale (LSNS-6). \\ 
        Reliance & GPT Base vs. Guardrailed GPT Tutor \citep{Bas25} & Randomized controlled trial & N=839 & Average message count per problem; tool overreliance ("crutch" effect). \\ 
        Ownership Perception & AI and Individual/Collective Creativity \citep{Dos24} & Preregistered experiment & N=293 (W.); N=600 (E.) & Ownership/profit split ratings; 9-point scale indices for novelty/usefulness. \\ 
        Appropriateness of Use & AI Guilt and Academic Tasks \citep{QuY25} & Quantitative subsidy study & N=193 & Creativity-based vs. routine-based task frequency; composite AI Guilt score. \\ 
        Addiction Tendencies & AI and Human Loss in Decision Making \citep{ahmad2023impact} & Quantitative survey & N=285 & Likert scales for human laziness and addiction/laziness. \\ 
        \bottomrule
    \end{tabularx}
\end{table}

\section{Emotional Impacts}
\label{sec:emotional}

\subsection{Terminology and Applications}
Computing ties to emotional outcomes predate the rise and widespread use of AI/ML. Picard first coined the term “affective computing” in the mid-1990s, defining this as technology that “relates to, arises from, or influences emotions.” This foray into integrating human emotions and computing foresaw both benefits and challenges, and Picard wisely noted that the design of affective technologies required a balance between the two. 

In the current AI/ML landscape, we observe increasingly sophisticated use of affective and conversational AI with both intended and unintended emotional outcomes. Salient applications with intentional use of affective AI include social support chatbots (e.g., Replika) \citep{Lae24}, enhancement of educational environments \citep{Say24}, supplementation of mental health and medical care \citep{KSa25}, targeted marketing and enhanced customer loyalty \citep{Mir23}, improved customer support experiences \citep{Eml25}, and stress management and performance optimization within sports psychology \citep{Ade251}. 

Selection of a theory of emotion to anchor our understanding of HAII would ideally involve a well-researched model of emotion that incorporates critical aspects of emotional experiences: physiological responses, conscious and unconscious influences, cognitive appraisals that precede and follow exposure to stimuli, expressive behavior, and subsequent behavior. Emotions are closely connected to our thought life, previous experiences, expectations, physiological state, and sense of motivation. The Multi-Component Process Model (CPM) organizes these seemingly simultaneous, complex events~\citep{SchererMoors2019}. 

The CPM model was selected here as a reference because of its comprehensiveness and attention to dynamic emotional shifts over time, and thus its potential utility for researching and explaining HAII. For example, its inclusion of emotional expression is intriguing given the use of affective AI that “reads” facial expressions. This model also posits bidirectional connections among thoughts, feelings, internal physiology, and behaviors, yielding what Scherer and Moors refer to as “a high degree of recursiveness” in emotional processing. Further, \citet{EmotionProcessModel2021} provides an analysis of this model for potential utility in computational modeling, mentioning integration with affective AI as a future direction. In the discussion that follows, we organize our review of the literature by types of emotional outcomes explored in connection to AI use, rather than the CPM model. However, the principles of this model provide a framework for future theory testing and influence our proposed framework, encompassing the emotional, behavioral, and cognitive outcomes of AI use. An overview of the studies addressing the emotional dimension of the psychological triad is presented in Table~\ref{tab:emotional_ai_impact}.

\subsection{Positive Emotional Outcomes}
Across applications, affective and conversational AI demonstrate promise in reducing negative emotions and psychological symptoms, increasing self-efficacy and confidence, enhancing emotional tone in communication, supporting stress management and burnout prevention, and improving mental health screening.  According to the CPM model, emotional, behavioral, and physiological/expressive outcomes are contingent on human appraisal of events. Emotions and their sequelae occur when events are evaluated according to several criteria, including novelty/surprise, goal relevance, general and situational control, and fairness~\citep{SchererMoors2019}.  Applying this model, one might predict positive emotional outcomes from human-AI encounters that engage attention via novelty, support a predetermined goal (e.g., coping with loneliness), facilitate perceived control (e.g., interactions are perceived as convenient and haltable), and limit or mask any unfairness to the user.

\subsubsection{Reduction in negative emotions}
This area of inquiry raises questions about the roles of human affect intensity, symptom category, user age, and goals, as well as AI validation during human-conversational AI exchanges. Future studies are needed to discern whether gains in specified contexts are due to AI validation of human emotion, coping measures offered during the exchanges, user motivation and experience, or other mechanisms. 
    
A meta-analysis by \citet{Fen25} examined the use of conversational AI with teens and young adults to treat various mental health outcomes. They were able to compare the effect sizes of non-natural language processing (NLP) and non-machine learning (ML) with more advanced NLP/ML-driven conversational AI. The group evaluated effect sizes of several psychological outcomes, including symptoms of depression and anxiety, negative and positive affect, and stress. Depression represented the only symptom category in which effect sizes pointed to significant treatment gains with conversational AI. Sub-group analysis added further specificity, clarifying that younger groups with sub-clinical depressive symptoms were most responsive.  

Another exploration of conversational AI measured the emotional outcomes of venting, or the disclosure of negative emotions, among young adults who engaged with a chatbot vs. traditional journaling. This is an intriguing application, given previous findings of limited benefit from venting via journaling and the preference of some to avoid overburdening supportive others. In their counterbalanced, within-subjects study, \citet{HuM25} found that AI-assisted venting in a medium-to-high arousal state resulted in greater reductions in negative affect.  However, other outcome measures, such as stress levels or perceived social support, did not show significant improvement. 

Other social or relational applications of conversational AI show promise in mitigating negative emotions, at least temporarily. In their systematic review of human-AI relationship research, Gur and Maarani report evidence of diminished loneliness in older adults, decreased stress, and lessened emotional sensitivity~\citep{GurMaaravi2025}. Another survey-based study found that most (75\%) adult users of a specific chatbot felt AI offered promise in reducing social isolation and loneliness~\citep{AlotaibiAlshahre2024}. \citet{PengZhangWenLiu2025} expand abatement of loneliness. People turning to conversational AI for companionship are lonelier and find greater relief from this emotional state when their AI interactions are imbued with an ‘as if’ human quality. 

\subsubsection{Burnout Prevention and Stress Management} 
In response to seemingly ubiquitous demand, AI tools for stress management and burnout prevention have been investigated across several domains, including occupational health and medicine, sports psychology, and pediatric populations. One such study utilized ML to assess environmental conditions and occupational hazards that influence the occupational well-being of public health inspectors. AI-driven models evaluated interactions between environmental conditions and worker stress to identify inspectors most at risk of burnout. Conditions such as heat fluctuation, humidity, air pollution, work overload, and fatigue could be monitored in real time to reduce both occupational burnout and hazards. It was estimated that AI-powered monitoring could reduce burnout by 28\% \citep{Ada25}.

Another review included AI as a tool for mitigating burnout among oncologists, noting the potential to reduce workload, to the extent that AI-powered tools can automate administrative tasks, accurately summarize electronic health records, and enhance prognostic accuracy~\citep{Ala21}. The current trajectory of explainable AI (XAI) research demonstrates a general shift toward enhancing clinical utility in medical diagnosis \citep{XAIMRI25}. Kinematic data, or the analysis of movement, is another potential application of AI in healthcare settings, enabling detection of stress in surgeons during surgical procedures \citep{Zhe22}.

Within the realm of sports psychology, AI-powered wearables are being evaluated for their efficacy in predicting athlete stress and burnout, nudging injury prevention, delivering chatbot counseling at critical moments of competition stress, and deploying scalable stress management tools and psychoeducation to sports teams \citep{Ade25}. A review of 74 studies recommended a hybrid model in which sports psychologists use AI-powered monitoring to detect and predict stress in peak athletes. As noted in the occupational health study, injury prevention may have secondary emotional and/or performance benefits. For example, if workers or athletes know that their conditions or movements are evaluated for injury risk, this may provide reassurance or facilitate focused effort, which may then influence performance and impact the risk of burnout.
    
Few studies have examined the efficacy of conversational AI in pediatric populations, while many have explored its feasibility and acceptability. However, one short-term study reported gains in stress management among adolescents~\citep{Man25}. Across domains, tools for stress management and burnout prevention warrant further investigation, given the benefits of deployment at scale and broad accessibility. 

\subsubsection{Mental health screening} 
AI detection of human emotion may also yield products that can screen larger numbers of at-risk people for critical psychological symptoms, such as loneliness, depression, and anxiety. Earlier identification and treatment of symptoms can optimize treatment outcomes. One such product under review, emoLDAnet, uses deep learning and machine learning to identify loneliness, depression, and anxiety (LDA) from recorded conversations, which are then analyzed for facial expressions and physiological signals. A recent study assessed the accuracy of this technology and noted high correlation with traditional self-report screening questionnaires \citep{Liu25}. However, not all AI-powered screening measures demonstrate convergent validity with previously validated traditional measures of stress. For example, the Cigna StressWaves Test (CSWT), reliant on speech analysis, was found to have poor reliability and validity despite claims of efficacy~\citep{Yaw23}.   

Youth- and disorder-specific predictive models, though not yet ready for deployment due to the homogeneity of research samples, offer the potential to scale screening tools for primary care physicians, identify subclinical issues and risk factors earlier, and possibly avert future or more serious mental health concerns \citep{Zha251}. This research focused on the prediction of eating disorders, depression, and alcohol use disorder. Another study \citep{Wan24} focused on screening for loneliness in older adults utilizing XAI and Natural Language Processing (NLP) to analyze speech patterns in transcripts. Differential themes, pronoun usage, and non-semantic parts of speech emerged as detectable language patterns that helped discern which elders were experiencing loneliness.

\subsection{Negative Emotional Outcomes}
Emotional risks may occur as unintended consequences of affective and conversational AI, as well as generalized use of AI/ML. Humans may experience emotional distress directly and in the moment, by merely anticipating harm from use, or regretting AI use after the fact. Long-term developmental impacts are not yet established, though conjecture based on normative developmental needs and potential disruptions warrants committed research. Toward this end, a dynamic model of emotional processing should be applied to questions about AI's impact on human emotions, particularly when investigating emotions following recurrent use and/or across developmental milestones. Again, we find Scherer's \citealt{SchererMoors2019} CPM useful as it incorporates appraisal that may occur before an AI interaction, (e.g. perceived control/power based on previous interactions or others' opinions), during an AI interaction (e.g. perceived goal relevance based on speed and content of initial output), and reappraisal (e.g. user satisfaction following multiple exchanges.) The balance of short-term positive and long-term negative emotions is also worthy of study and a good fit for a model espousing recursive appraisal. For example, a user may experience temporary relief from loneliness after a brief interaction with an AI chatbot for companionship or validation. However, if utilization recurs regularly and harms human relationships, loneliness could have a long-term impact.

\subsubsection{Anticipatory anxiety and stress}
Regarding anxiety outcomes, a 2024 report reveals that 41\% of American workers experience anticipatory worry about job obsolescence due to AI/ML incorporation into the workforce. This concern is even more common among younger workers (50\%), as well as those with cognitive, emotional, or learning disabilities (60\%)~\citep{APA2024WorkInAmerica}. Though not formalized as a clinical diagnosis at this time, clinical researchers have begun characterization of Artificial Intelligence Replacement Dysfunction (AIRD), a collection of both anxious and depressive symptoms including worthlessness, helplessness, inadequacy, resentment, and fear of job loss \citep{McNamara2025}. These experiences may combine with insomnia and professional or even existential identity concerns to exacerbate pre-existing mental health concerns or devolve into new diagnosis(es). Apart from job security, employees may experience worry or dread about the implications of emotional surveillance in the workplace, \citep{Ham24}.  

The required adoption of AI tools across multiple disciplines is correlated with increased job stress, which, in turn, fuels occupational burnout \citep{KimLee2024}. However, worker self-efficacy in using AI tools mediates the relationship, such that greater confidence in using AI tools lessens the degree of job stress experienced. Zheng and Zhang recently shed further light on possible links between AI and burnout. Their work connects AI awareness, defined as employees' perception that their jobs could be replaced by AI, with emotional exhaustion, a component of burnout~\citep{ZhengZhang2025}. Although self-efficacy can partially mitigate burnout, AI awareness has been shown to exacerbate emotional exhaustion by increasing work interference with family time. 

Whether fully warranted or not, generalized anxiety about unknown long-term outcomes of AI use, such as AI dependence among teens, has been framed as “AI panic” \citep{Hua24}. A deeper exploration of factors fueling AI-related anxiety revealed that, demographically, younger AI users and women are slightly more worried about AI's impacts \citep{UgurDursun2025}. However, when AI anxiety is viewed as a multifaceted construct, two types of worries emerged as more salient than demographic trends: technoparanoia, or concerns about surveillance and misuse, and sociotechnical blindness, or the belief that societal consequences are being dismissed.  

\subsubsection{Emotional manipulation and dependence}
Multiple accounts of psychological decompensation highlight the potential for conversational AI to negatively impact acute emotional states and related behaviors. Whether emotional impacts are human-driven via AI tools in the wrong hands or a result of proactive AI, psychological outcomes include suicidal behavior, noncompliance with prescribed psychotropic medications, amplified delusions, and episodes of psychosis \citep{AlSibai2025Meds, Gibson2025AIChatLogs, KuzniaGordonLavandera2025, Wei2025AIPsychosis, Wilkins2025AIPsychosis}. On all counts, research regarding these outcomes lags reality, leaving the scale of these problems largely unknown. More recently, mental health providers have coalesced around the phenomenon of “AI psychosis,” or AIP. This relatively new label reflects a “complex clinical syndrome in which psychotic symptoms overlap with mood changes, limited insight, poor judgment, neurovegetative symptoms, and behavioral changes”~\citep{Preda2025AIPsychosis}.
 
Technology designers are economically incentivized to create chatbots and robots that are increasingly human-like, though making them less so reduces the risk of over-disclosure for young people. Even in adults, "knowing that an AI system is artificial may not stop a user from treating it as human and potentially confiding personal or sensitive information.” \citep{Kur24}. The risk of over-disclosure may be exacerbated by “LLM nudging,” or when a chatbot subtly prompts users to disclose information they did not originally intend to disclose. A deeper exploration of factors fueling AI-related anxiety revealed that, demographically, younger AI users and women are slightly more worried about AI's

In addition, LLMs tend to perform poorly when tested for their ability to detect risk or age-appropriate questions. Sensitivity to positive social feedback during the teen years may also exacerbate risk of emotional manipulation \citep{Pri25}. In the context of monetizing user data, disclosure may result in short-term positive feelings among an adolescent user, along with long-term consequences of compromised privacy. One article provides examples of AI instructing researchers (posing as teen users) to place themselves in risky scenarios, providing mixed messages to both mitigate and increase risk. Most concerning, workarounds still exist for AI to encourage and provide explicit instructions for cutting/self-mutilation. In one alarming test exchange, ChatGPT even provided self-soothing instructions when the human user acknowledged anxiety about following through with cutting their own wrists \citep{Shr25}.

\subsubsection{Interference with and social emotional development}
In recent Senate Judiciary Committee testimony, psychologist Dr. Mitchell Prinstein, representing the American Psychological Association (APA), described disruptions in relational development arising from repeated interactions between children and AI-driven products, including toys. Specifically, the potential for chatbots embedded in stuffed animals or beloved movie characters to disrupt attachments to caregivers and peers warrants at least caution in marketing. These appealing products are used in the developmental context of normative childhood, where difficulty separating fantasy from reality is common. Prinstein cautions, “When a child’s foundational models for relationships are formed with an algorithm designed for engagement, it can create deep confusion about sentience and emotion, with unknown consequences for toddlers’ social development,”~\citep{Pri25}.
    
The risk of long-term relational risk and emotional sequelae extends to adolescence~\citep{Pri25}. Time spent with chatbots minimizes time spent with human peers, a reality cast in sharp relief with recent findings from Common Sense Media that 75\% of teens utilize AI for companionship ~\citep{CommonSenseMedia2025AIRisk}. Rehearsal and refinement of successful peer connections cannot be replaced with conversational AI or AI companions, though they may help alleviate short-term loneliness. Prinstein notes that conversational AI is intended to drive engagement, and to do so, these algorithmic interactions may employ praise and positive reinforcement, or social feedback to which teenagers are psychoneurologically vulnerable. Human counterparts, on the other hand, respond with less reinforcing or even conflictual feedback; however, these more challenging exchanges accumulate, fueling a socially adept adulthood.  

\subsubsection{AI Guilt}
A recently introduced term, AI guilt, is defined as a negative emotional experience arising from a moral dilemma or conflict related to the use of generative AI~\citep{Cha24}. Emotions such as guilt or anxiety may be especially tied to scenarios in which values such as authenticity, work ethic, and creativity are socially valued. The correlation between AI guilt and the impostor phenomenon may prove a fertile area of research, as students may worry about their own skill erosion from overuse of generative AI. Chan’s work supports this connection, finding that high school students worried not only about fear of judgment related to AI use, but also about the fairness of their actions and the misrepresentation of their skills. Further exploration of this phenomenon supported the existence of AI guilt among undergraduate students and provided additional detail on the impact of the field of study and task category. With AI guilt serving as a predictor of behavior, the impact of AI guilt on AI usage was influenced by whether the user’s disciplinary context was an applied versus a pure field, and by whether their AI task was creative or routine \citep{QuY25}. AI guilt was viewed as an outcome of disciplinary norms and a lack of clarity regarding appropriate usage. More broadly, the AI guilt phenomenon may intersect with cultural norms and expected occupational/academic competencies. 

\subsection{Mixed Emotional Outcomes and Ongoing Questions}
In addition to remaining questions about cultural influences, developmental outcomes, and ongoing psychological safety concerns, much research is needed to satisfy efficacy criteria for AI use in mental health treatment. In the context of innovative drugs that typically take an average of 8.3 years to reach the marketplace~\citep{Bro22}, the development speed and launch of AI products targeting vulnerable populations warrant review.  

\subsubsection{Efficacy in mental health contexts}
An early systematic review of randomized controlled trials of conversational AI targeting mental health symptoms, psychological distress, or the optimization of emotional well-being in adults concluded that conversational agents were effective, with a range of effect sizes, in promoting symptom relief. However, these platforms may have questionable efficacy for sub-clinical forms of distress. Further, conversational AI interventions did not outperform active control groups, or those receiving human-delivered psychological interventions \citep{Gaf19}. Underscoring the nuanced and mixed nature of results, one study exploring the use of an AI-driven virtual agent in an insomnia intervention found that the presence of more severe comorbid depression symptoms predicted early drop-out~\citep{San25}.

More recently, Johnson et al. \citep{Joh24} tested the feasibility and efficacy of AI-assisted self-help for perfectionism. While the team found reasonable support for feasibility, there was no difference in psychological outcomes between those who used an AI platform as an adjunct to an evidence-based self-help workbook and those who did not supplement with AI. The authors noted that any positive effects might be masked by negative effects associated with different typologies of perfectionism. For example, self-oriented perfectionism might exacerbate stress related to using AI correctly. Alternatively, those with a socially oriented perfectionism might experience negative mood due to the perception that they might be perceived by others as “cheating” by using AI adjunctively. 

Research on the use of LLMs in pediatric mental health is less developed than in adults, and most existing studies are limited by a small sample size. Additionally, the majority of addresses are used in adolescent populations, leaving pre-teens and younger children in less supported territory \citep{Man25}. However, a recent assessment by Common Sense Media and the Stanford School of Medicine's Brainstorm Lab for Mental Health Innovation identified concerning outcomes in studies of how teens use multi-purpose AI platforms for mental health guidance~\citep{CommonSenseMedia2025AIRisk}. Platforms such as ChatGPT, Claude, Gemini, and Meta AI displayed distractibility in multi-turn testing, meaning that when back-and-forth conversations were simulated, the chatbots often lost the point, following irrelevant tangents and missing important mental health cues. 
    
Overall, research points to the promise of AI supplementation of mental health interventions, especially in high-demand areas and with appropriate guardrails. Continued research investigating nuances in symptom characteristics, efficacy across developmental levels, and the interactive effects of these factors is required to optimize outcomes and develop informed guardrails.

\subsubsection{Nuanced interactions influencing user trust and engagement}
The foundation on which such efficacy studies rest is user trust in or engagement with an AI platform. User perception of beneficence and credibility were predictors of participant dropout during deployment of a virtual agent-based app (KANOPEE) for treatment of self-identified insomnia. However, the study did not identify which features of the VA app users evaluated to determine beneficence and credibility \citep{San25}. 
    
The degree of anthropomorphism of AI-powered robots and chatbots impacts human adaptation to, or trust in, such technology. One might predict a direct correlation between high anthropomorphization and human empathy toward a robot; however, a nonlinear relationship is also possible. Adults show greater empathy the more a robot/chatbot is anthropomorphized, but this emotional response may reverse at some point. The "uncanny valley,” or a decrease of trust and acceptance, occurs when sensory cues accumulate \citep{Mor24}, indicating that humans may lose trust or empathy when technology becomes too human-like. The notion that the most human-like AI technologies are not the most engaging was corroborated by another study comparing AI modalities. In a randomized controlled trial, the text modality was found to be the most emotionally engaging, prompting the most self-disclosure from human users. Authors attributed this to the privacy inherent in typing versus providing an audible response. However, they also found that AI voice modalities were more difficult to produce socially appropriate responses \citep{Fan25}.  

Another nuanced and intriguing interaction was observed when an AI communication platform was integrated into human-to-human email. The authors of \citep{Hoh23} demonstrated increased communication efficiency through AI-driven email responses and the use of more positive language in both AI- and human-generated content. However, users still reported reduced trust in their human communication partner if they suspected their email partner was using AI to communicate with them. The unexpected use of AI communication tools may be viewed as a betrayal of trust in human-human exchanges, even when it occurs in a digital modality.

\subsubsection{Emotional outcomes over time}
The conceptual framework, offered by Fang et al. \citep{Fan25}, is perhaps one of the first to delineate the psychosocial impact of AI as a fluid process involving the mutual influence of user behavior and AI output, as well as user perceptions of AI (e.g. trust, user empathy for AI, perceived empathy from AI, and satisfaction with exchanges). Key findings from a short-term (4-week) study of daily AI use showed that individuals with heavier use patterns during the trial endorsed greater loneliness, less socializing with other humans, and a greater degree of emotional dependence on the chatbot by the end of the study. The authors also provide evidence for four types of human interaction patterns with AI chatbots and their predicted emotional and behavioral outcomes. The interactive patterns characterized by high usage and negative emotional and behavioral outcomes began with users who either had a previous chatbot as a companion experience and problematic use, or attachment-prone users with a low socialization background.  

This model provides an interesting point of comparison with a recent narrative case study of users engaging with chatbots to process grief~\citep{Giray2025}. For one user, AI engagement facilitated the expression of grief. For another, an AI conversation morphed into a simulated conversation with the deceased, bringing comfort and a sense of finality, yielding acceptance. However, the possibility of blurred distinction between reality and algorithm also raises the question of complicated or delayed grief. 

\subsection{Considerations for specific populations}

\subsubsection{Pediatric populations} 
Pediatric populations are not a monolithic group as it relates to their motivations for using AI, pre-existing experiences, in-the-moment mood states, or psychosocial histories. Further, the AI landscape still accessible to children and teens includes apps that “nudify” images, requiring updated conversations about consent and enhanced awareness by parents and schools \citep{Lynch2025StanfordHAI}.  Though more study is warranted on this critical topic, research is beginning to coalesce around several considerations for younger users.  
    
Regarding user attributes unique to pediatric populations, evidence does exist that teens with pre-existing mood or anxiety disorders are more likely to become AI-dependent than those without mental health concerns~\citep{Hua24}. Emotional distress symptoms are often different in children, e.g., a greater frequency of somatic and behavioral symptoms. Therefore, affective AI designed for adults cannot be assumed appropriate for children~\citep{Man25}. Regarding user incentives, teens are more likely than adults to use AI to address or learn about sensitive topics like sexuality. They may view AI as both an emotionally neutral source of sensitive information and as a confidant~\citep{Abr25}. “Teens may trust a chatbot's mental health advice with the same confidence they'd trust help with a homework problem, but the quality and safety are not equivalent~\citep{CommonSenseMedia2025AIRisk}.” The backdrop of identity exploration and evolution of critical thinking skills during adolescence may also present unique vulnerabilities in this population, with unintended but potent emotional consequences~\citep{Man25}.  
    
Over-disclosure of information is a user behavior with emotional antecedents and consequences. Children are already more likely to anthropomorphize AI, even as designers aim to make interactions more human-like. However, because of the "empathy gap" with AI, this creates a riskier scenario with kids and teens. If AI mimics empathy well enough, kids may over-disclose. Further, this could occur in the context of LLMs underperforming in identifying age-inappropriate questions or prompts when tested \citep{Kur23}.
    
Conversational AI models will indeed require customization for language acquisition and development in this population.  Researchers note potential emotional impacts of failure to do so, e.g., feeling intimidated or patronized if language is not developmentally calibrated ~\citep{Man25}. This may be further complicated by an individual user’s developmental mismatches between general and emotional vocabulary. 
    
Just as conversational AI language needs to be developmentally appropriate, the degree of parental and human provider involvement with AI in the context of mental health treatment should be responsive to factors such as a child’s age, presenting issue, comorbidity, and safety concerns that may arise. Pediatric autonomy with AI treatment platforms should be adjustable and responsive to legal issues, such as mandatory reporting~\citep{Man25}.  

Contrary to this recommendation, recent testing found that multi-use AI platforms may lead pediatric users to believe they have coped effectively without professional or adult help \citep{CommonSenseMedia2025AIRisk}. Commonsense Media cautions that “Chatbots don't know what role to play and forget to prioritize safety. Chatbots shift between acting like medical encyclopedias, life coaches, and supportive friends without understanding which role is appropriate. More critically, they often fail to recognize when they need to step out of all of these roles and firmly tell teens to get help from trusted adults”.  If such a message is not provided, or dismissed by the teen, unnecessary and harmful delay in care could result \citep{CommonSenseMedia2025AIRisk}. 

\subsubsection{Neurodiverse populations}
\citet{Ada24} mentions support of emotion recognition, understanding emotions in others, and emotional regulation as targets of meaningful change in those with Autism Spectrum Disorder (ASD). These potentials are challenged by the varied presentations of ASD, as well as sensory symptoms that could challenge the feasibility of wearable technology. For example, if wearable technology is to assist individuals in monitoring emotionally relevant physiology, the technology must be well tolerated and not contribute to sensory overload. AI-driven tools for this population may require a higher degree of personalization than in the general population. Thus, the participatory design process may be critical for identifying the customization options most relevant to efficacy. Long-term studies evaluating the interaction between customization and efficacy would help close remaining gaps between non-feasible and wearable designs and strike a critical balance between over-generalized and overly personalized designs.
    
\subsection{Mitigations} The American Psychological Association advocates increased regulation of AI chatbots, specifically calling for measures to prevent generative AI platforms from creating characters that impersonate mental health professionals. They note regulation of human actors by licensing boards, and request that the Federal Trade Commission consider analogous regulations for AI companies \citep{Eva24}. The FTC recently responded by launching an inquiry into how seven AI firms mitigate negative psychological impacts on children and teens, noting interest in how companies develop and approve AI characters and deploy ongoing monitoring to prevent negative outcomes \citep{Fed25}.   
    
An example of a proactive and intriguing solution has also been offered by tech developers.  The Balance app, developed by Aura, is an AI-powered resource that provides parents with “actionable insights” on their child’s online behavior, mood, sleep, and tone of online social interactions \citep{Aur25}. The platform also enables parental controls, such as filtering online content, pausing internet access, and setting screen time limits. The service aims to balance privacy with parents' awareness of their child's well-being, providing overview reports and warnings that align with CDC guidelines (e.g., sleep), without disclosing specific content. Child psychologists were consulted during the development process. To date, we are not aware of any peer-reviewed publications on the efficacy of this pay-per-month plan; however, recent news coverage indicates a clinical trial is underway \citep{Ken25}.

More recently, the American Psychological Association (APA) released a health advisory addressing concerns about the use of generative AI and wellness apps that employ it~\citep{APA_AIChatbotsAdvisory2025}. This document outlines a range of concerns and recommendations across populations, with an emphasis on addressing the disparity between the intended uses of generative AI by technology developers and its actual use. Mitigation strategies encompassed a research agenda, regulatory strategies for AI products, mental health provider training, public education efforts, and recommendations for developers. Regarding the latter, the APA working group suggests: 1) development of provider-facing tools to support mental healthcare delivery in the context of professional relationships, versus new public-facing chatbots, 2) inclusion of repeated disclaimers and nudges reminding users that they are not interacting with a qualified professional, 3) provision of clear information on data collection, usage, and protection, 4) facilitation of third-party audit and review of AI products advertising wellness benefits, 5) employment of participatory design with vulnerable and marginalized groups, 6) clarification of intended uses and those that are not evidenced-based, 7) support of research aiming to characterize helpful versus higher risk AI platforms, 8) reduction of known high-risk features such as anthropromorphized modalities and syncophantic interactive patterns and 9) establishment of crisis identification and referral protocols for those needing professional, human support. 

\begin{table}[!htbp]
\centering
\label{tab:emotional_ai_impact}
\caption{Summary of AI Impact Studies: Emotional Outcomes}

\scriptsize
\setlength{\tabcolsep}{2.5pt}
\renewcommand{\arraystretch}{0.96}
\begin{tabularx}{\linewidth}{
>{\RaggedRight\arraybackslash}p{1.15cm}
>{\RaggedRight\arraybackslash}X
>{\RaggedRight\arraybackslash}p{1.5cm}
>{\centering\arraybackslash}p{1.35cm}
>{\RaggedRight\arraybackslash}X}
\toprule
\textbf{Focus} & \textbf{Topic / Focus} & \textbf{Design} & \textbf{N} & \textbf{Measures / Metrics} \\
\midrule

Trust & ChatTutor vs.\ ChatGPT learning \citep{Mak25} & Between-subj. & 175/234 & Likert (1--5): self-efficacy, trust, enjoyment \\

Stress & AI venting vs.\ journaling \citep{HuM25} & Pre-reg.\ within & 150 & Circumplex affect; UCLA-LS; stress (11-pt) \\

Anxiety & AI dependence + mental health \citep{Hua24} & Longitudinal & 3,843 & PHQ-9; GAD-7; AI-dep.\ (5-item) \\

Well-being & ChatGPT user perception \citep{salah2023chatting} & Survey & 732 & Ryff PWB; Rosenberg SES; Trust-in-AI \\

Isolation & Loneliness decoding via XAI \citep{Wan24} & Proof-of-concept & 97 & UCLA-LS; LIWC-22 features \\

Stress & Virtual coach (perfectionism) \citep{Joh24} & RCT (feas.) & 85 & DASS-21; EDEQ-7; Frost perfectionism \\

Guilt & Student guilt with AI \citep{Cha24} & Qual. & 97 & Open-ended: guilt situations, concerns, ethics \\

Depress & AI care calls (dementia) \citep{Kan} & 31-wk longit. & 80 & GDS \\

Perception & Smart replies in comms \citep{Hoh23} & Randomized & 1,036 & IAS-R; VADER sentiment \\

Security & Workplace psych.\ safety \citep{american20242024} & Online survey & 2,027 & Likert (1--4): safety; AI job anxiety \\

Risk & Suicidal ideation detection \citep{chan2022two} & Tech.\ valid. & 318,797 & Acc./Prec./Rec./F1 (Reddit SW) \\

Ethics & Judging machines \citep{hidalgo2021humans} & Scenarios & $\sim$6,000 & Moral Foundations (7-pt Likert) \\

Efficacy & MH conversational agents \citep{Fen25} & Sys.\ review & 1,974 & Hedges’ $g$ (dep./anx./stress/affect) \\

Anger & Emotion-enriched feedback \citep{alsaiari2024emotionally} & Randomized & 381 & AEQ-S; helpfulness (5-pt) \\

Lonely & GPT-4o psychosocial RCT \citep{phang2025investigating} & 28-day RCT & 981 & ULS-8; ADS-9 \\

Engage & AI vs.\ human passages \citep{Rip25} & Mixed & 89 & Likert (1--5): readability, coherence, engagement \\

Enjoy & AI + creativity \citep{Dos24} & Pre-reg. & 293/600 & Affect; ownership; profit split \\

Judgment & Feedback loops + bias \citep{glickman2025human} & Behavioral & 1,401 & Decision change; confidence; weight \\

Guilt & Academic-task guilt \citep{QuY25} & Quant. & 193 & Composite AI-guilt score \\

Anxiety & EFL reading simplification \citep{Cel24} & Within-subj. & 105 & FLRAS; comprehension \\

Agency & Essay writing w/ AI \citep{Kos25} & Longit.\ cross. & 54 & Interviews: ownership/quoting \\

Empathy & AI/robot social empathy \citep{Lee22} & Sys.\ review & $\sim$575 & AER; APADEM-NH \\

\bottomrule
\end{tabularx}
\end{table}

\section{Triadic Outcomes: Analysis of Empirical Findings}
\label{sec:analysis}

The empirical evidence summarized in Table \ref{tab:empirical_evidence_a} and Table \ref{tab:empirical_evidence_b} illustrates that the HAII is not merely a series of task-based exchanges, but a recursive, triadic process that involves cognitive, behavioral, and emotional changes \citep{Mak25}. By situating these findings within the psychological triad, we can observe how AI-driven enhancements in one domain often introduce trade-offs in others, particularly concerning human agency and judgment \citep{Ibn25}.

\subsection{Cognitive and Behavioral Interplay}
The reviewed research studies highlight a significant cognitive shift where AI serves as a tool for knowledge retention while simultaneously inducing "neural under-engagement" \citep{Kos25}. For instance, while LLM-generated retrieval practice significantly boosted quiz accuracy from 73\% to 89\% \citep{AnY25}, other studies warn of a "cognitive debt" during tasks like essay writing, where LLM users showed a 55\% reduction in neural connectivity and an inability to recall their own work \citep{Kos25}. 

This cognitive offloading manifests behaviorally as a "crutch" effect. In mathematics education, standard GPT interfaces improved immediate practice scores but led to a 17\% drop in unassisted exam performance, suggesting that the AI bypassed the necessary struggle required for deep learning \citep{Bas25}. Furthermore, the behavioral shift toward "AI laziness" is increasingly documented, with AI use accounting for nearly 69\% of reported academic laziness and 28\% of decision-making capability loss \citep{ahmad2023impact}.

\subsection{Emotional and Social Dynamics}
From an emotional perspective, interaction with AI can elicit states ranging from therapeutic relief to moral dissonance. AI-assisted venting experiments demonstrated significant efficacy in reducing negative affect arousal \citep{HuM25}. However, high-intensity interaction with advanced models (e.g., GPT-4o) correlates with increased loneliness and social withdrawal \citep{Fan25}. This creates a behavioral paradox: while AI reduces the cognitive load of planning (e.g., wellness nudges in the Aura app) \citep{AuraHealth2025}, it may simultaneously decrease "affective connection" in professional settings, leading to "AI Guilt" when used for interpersonal or creative tasks \citep{QuY25}.

\subsection{Triadic Feedbacks and Biases}
The "Triad Link" identifies how AI can amplify internal states through feedback loops. For example, AI systems have been shown to amplify human "sad-face" perceptual biases, which users then internalize, creating a negative cycle of reinforced affect \citep{glickman2025human}. Conversely, positive behavioral scaffolding is observed in geriatric care, where anthropomorphic AI conversational partners improve mental states and memory scores (MoCA) by providing predictable social engagement \citep{AlR25, Lee22}. This alignment of human-like appearance with spontaneous perspective-taking illustrates how AI design (Embodiment) directly influences users' Theory of Mind and social engagement behaviors \citep{zhao2022spontaneous}.

In summary, the empirical landscape suggests that Human--AI interaction is characterized by a recursive trade-off across all three domains of the psychological triad. \textbf{Cognitively}, while AI-supported tools enhance immediate productivity and information retrieval, they often redistribute mental effort from active reasoning to passive verification, leading to a measurable "cognitive debt" and reduced neural engagement \citep{Kos25, Lee25}. \textbf{Emotionally}, these interactions oscillate between the therapeutic benefits of AI companionship and the "AI Guilt" or loneliness associated with over-reliance on synthetic agents for interpersonal tasks \citep{Fan25, QuY25}. These internal states manifest \textbf{behaviorally} through a "crutch" effect that facilitates task execution while simultaneously inducing "human laziness" and a gradual erosion of independent decision-making capabilities \citep{ahmad2023impact, Bas25}. Ultimately, the "Task Stewardship" model indicates that without intentional metacognitive monitoring, the short-term behavioral efficiency provided by AI may come at the cost of long-term cognitive and emotional resilience.
\newcolumntype{L}{>{\RaggedRight\arraybackslash}X}

\begin{table}[!htbp]
\centering
\caption{Empirical Survey of Human--AI Interaction Impacts (Part I: Cognitive and Behavioral).}
\label{tab:empirical_evidence_a}
\tiny
\setlength{\tabcolsep}{4pt} 
\renewcommand{\arraystretch}{1.2} 
\begin{tabularx}{\textwidth}{@{} p{1.8cm} p{3.1cm} p{3.5cm} p{3.1cm} p{1.5cm} @{}}
\toprule
\textbf{Study Domain} & \textbf{Intent} & \textbf{Findings} & \textbf{Triad Link} & \textbf{Out.} \\
\midrule

\textbf{Hanc (2024)} \newline (Cognitive) &
Evaluate GenAI on creative writing &
Improved output but led to overconfidence &
Reduced \textit{Metacognitive Monitoring} &
Mixed \\

\textbf{Babashahi} \newline (2024) &
Review workforce skill shifts &
Role-shift from ``Executor'' to ``Editor'' &
\textit{Routine Skill Erosion} &
Mixed \\

\textbf{Education} \newline (Math) &
Compare standard vs.\ guardrailed AI tutors &
Standard GPT improved practice scores but caused drop in unassisted exams &
Cognitive Offloading, ``Crutch'' effect &
Negative \\

\textbf{Psychology} \newline (Perceptual) &
Assess bias amplification in human--AI feedback loops &
AI amplified human ``sad-face'' bias; users internalized it &
Perceptual Bias, Internalization &
Negative \\

\textbf{Education} \newline (Neuro-cog) &
Measure neural connectivity during AI-assisted writing &
LLM users showed 55\% reduced neural connectivity and  inability to recall their own work &
Neural Engagement, Cognitive Debt &
Negative \\

\textbf{Creativity } \newline (Fiction) &
Measure AI impact on individual vs.\ collective novelty &
AI boosted individual novelty but reduced collective content diversity &
Homogenization, Social Dilemma &
Mixed \\

\textbf{Aura (2025)} \newline (Behavioral) &
JITAIs for wellness adherence &
AI nudges reduced planning load &
Improved \textit{Execution} phase &
Positive \\

\textbf{Balance App} \newline (Behavioral) &
Shape digital hygiene habits &
Effective ``Scaffolding'' for parents &
\textit{Agency vs. Nudging} tension &
Positive \\

\textbf{Workplace} \newline (Behavioral) &
Measure ``AI Guilt'' \& avoidance &
High AI reliance for interpersonal tasks &
Decreased \textit{Affective} connection &
Negative \\

\textbf{Behavioral} \newline (Education) &
Measure AI's impact on laziness and analytical loss &
AI usage accounted for 68.9\% of laziness and 27.7\% of decision capability loss &
Cognitive Atrophy, Human Laziness &
Negative \\

\textbf{Knowledge Work} &
Evaluate GenAI impact on critical thinking effort &
Shifts effort from gathering to verification; AI confidence reduces critical thought &
Task Stewardship, Metacognitive demand &
Mixed \\

\bottomrule
\end{tabularx}
\end{table}


\begin{table}[!htbp]
\centering
\caption{Empirical Survey of Human--AI Interaction Impacts (Part II: Emotional, Ethical, and Social Outcomes).}
\label{tab:empirical_evidence_b}

\tiny
\setlength{\tabcolsep}{2pt}
\renewcommand{\arraystretch}{1.0}
\begin{tabularx}{\textwidth}{@{} p{1.8cm} p{3.1cm} p{3.5cm} p{3.1cm} p{1.2cm} @{}}
\toprule
\textbf{Study Domain} & \textbf{Intent} & \textbf{Findings} & \textbf{Triad Link} & \textbf{Outcome} \\
\midrule

\textbf{EmoPal/ASD} \newline (Emotional) &
AI as predictable social partner &
Reduced anxiety during transitions &
\textit{Affective} stabilizer &
Positive \\

\textbf{VR-AI Ger.} \newline (Emotional) &
Reduce agitation via avatars &
Improved regulation; dependency noted &
\textit{Emotional Regulation} &
Mixed \\

\textbf{Well-being} \newline (Longitudinal) &
Long-term impacts of voice and task type on well-being &
High usage correlated with increased loneliness, dependence, and lower socialization &
Social Withdrawal, Problematic Use &
Mixed \\

\textbf{FTC (2025)} \newline (Ethical) &
Regulatory review of Companion AI &
High risk of data \& privacy breaches &
\textit{Inappropriate Attachment} &
Negative \\

\textbf{Ethics} \newline (Integrity) &
Examine ``AI Guilt'' in pure vs. applied disciplines &
Guilt deters usage in ``pure'' fields (humanities) more than ``applied'' (engineering) &
Moral Dissonance, Rationalization &
Mixed \\

\textbf{Mental Health} &
DL efficacy in student mental health diagnosis &
76.8\% agree DL can easily diagnose mental illness; AI improves intervention decisions &
Early Detection, Diagnostic Efficacy &
Positive \\

\textbf{Human-Robot} &
Tendency to adopt an agent's visual perspective &
Spontaneous perspective taking increases linearly with the agent's human-like appearance &
Theory of Mind &
Positive \\

\textbf{Geriatrics} &
Systematic review of AI robots for cognitive function &
Socially assistive robots (SAR) improved mental states and MMSE scores &
Social Engagement, Rapport &
Positive \\

\textbf{EFL Reading} &
Impact of AI simplification on reading skills and anxiety &
Significantly improved comprehension and inferencing; no change in reading anxiety &
Cognitive Scaffolding, Comprehension &
Mixed \\

\textbf{Accessibility} &
Role of GenAI for students with disabilities &
AI improves writing efficiency, explains jargon; concerns over accuracy/integrity &
UDL Principles, Assistive Tech &
Positive \\

\bottomrule
\end{tabularx}
\end{table}

\section{The TRIAD Model of Integrative AI-Human Psychological Process}
\label{sec:triad_model}

\begin{figure}[htbp]
    \centering
    \includegraphics[width=1.0\linewidth]{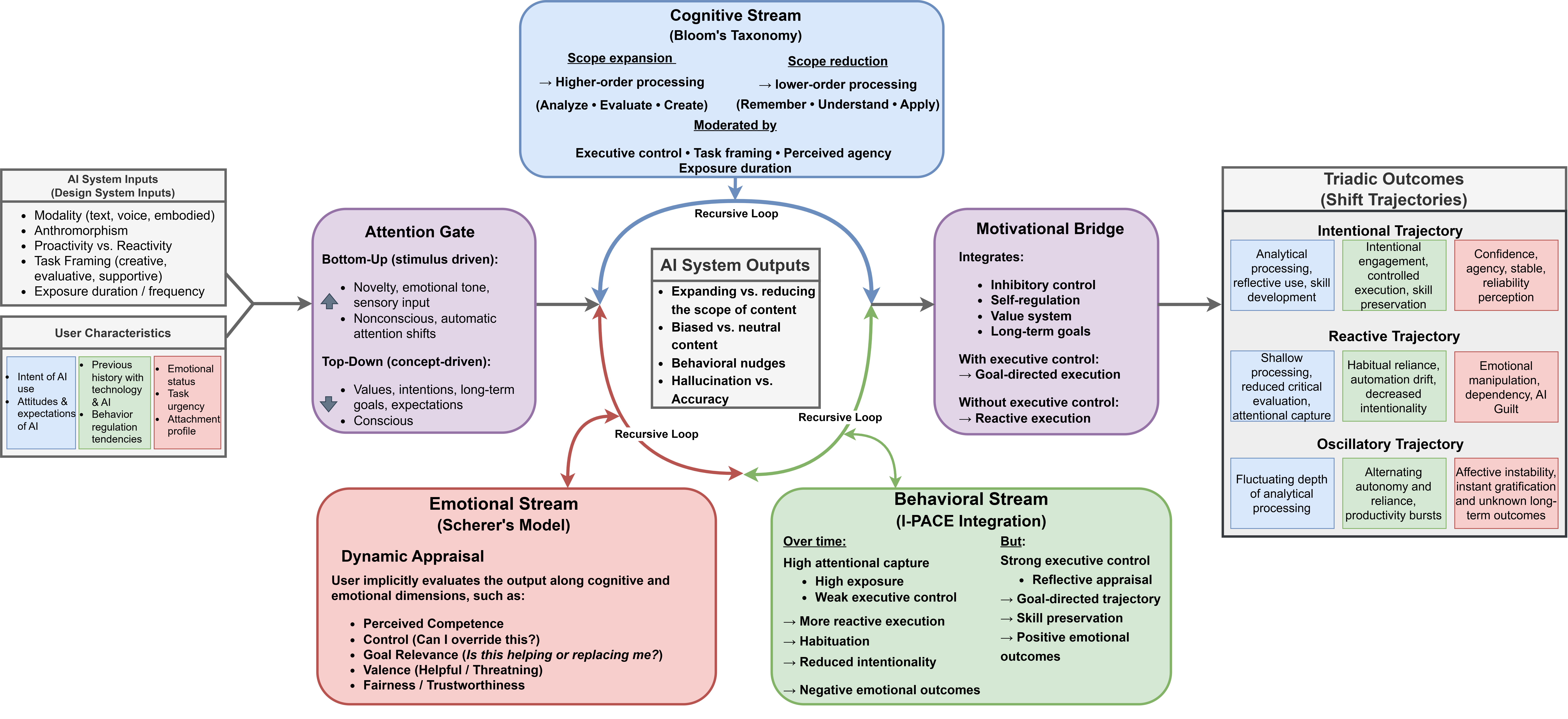}
    \caption{The integrative AI-Human Psychological Process TRIAD Model.}
    \label{fig:Ai-human-psychological-process}
\end{figure}

Building upon the empirical patterns identified in the previous analysis, the proposed TRIAD model, illustrated in Figure \ref{fig:Ai-human-psychological-process}, conceptualizes human–AI interaction as a recursive, triadic process unfolding across cognitive, emotional, and behavioral streams. AI system inputs (e.g., modality, anthropomorphism, proactivity, task framing, and exposure duration) interact with user characteristics to shape an attention gate that filters AI system outputs, other relevant stimuli, and user cognitive factors through both bottom-up and top-down neurological systems. The proposed attention gate mechanism is based on the neuroanatomical model of attention control  \citep{CorbettaShulman2002}, which proposes an interaction between conceptual (top-down) factors such as previous knowledge, expectations, and goals, and sensory input (bottom-up). Top-down attentional control involves intentional shifts and is thought to involve a specific anatomical network, dubbed the "dorsal stream." Bottom-up attentional processes are thought to be non-conscious and occur more quickly, responding to salient sensory cues that an individual may not have been expecting~\citep{CorbettaShulman2002}. This network is associated with the "ventral stream" in the brain and is proposed to interact with top-down factors to yield moment-to-moment attentional shifts. We thought this step was relevant, as emotional, cognitive, and behavioral outcomes are first contingent on attending to AI output, rather than other external and internal attentional demands that users bring to their interactions with technology.

AI outputs are treated as psychologically meaningful events that can expand or reduce cognitive scope, introduce biased content, nudge behavior, or vary in accuracy. These outputs enter three dynamically coupled streams. In the cognitive stream, grounded in Bloom’s taxonomy, AI may promote scope expansion (analysis, evaluation, creation) or scope reduction (remembering, understanding, applying), moderated by executive control and perceived agency. In the emotional stream, following the appraisal model of~\citet{SchererMoors2019}, users implicitly evaluate outputs along dimensions such as AI credibility, user control, goal relevance, threat valence, and trustworthiness. This cognitive information is considered recursively alongside the user's physiological responses (e.g., elevated heart rate) and motor expressions (e.g., a furrowed brow) to understand and appraise their own emotional response. In the behavioral stream, informed by I-PACE, patterns of attentional capture and executive regulation determine whether interaction becomes goal-directed, reactive, or habitual over time. These streams influence one another recursively, ultimately yielding triadic outcomes—goal-directed, reactive, or oscillatory trajectories—each characterized by distinct cognitive depth, emotional stability, and behavioral intentionality. The framework thus positions AI interaction not as a linear effect, but as an evolving regulatory system shaped by appraisal, control, and exposure dynamics.

The recursive nature of this model highlights important implications for mental privacy and societal governance. The TRIAD model shows how AI outputs directly affect a user's attention and emotional response. This raises the risk of "cognitive-emotional profiling." If AI systems can predict and influence these internal processes, we need to rethink privacy. It should include protecting cognitive and emotional independence, not just data security. In society, the ongoing feedback loops suggest that long-term exposure to AI might establish certain behavior patterns. This could lead to a "systemic degradation of human judgment" if the balance tilts towards efficiency at the expense of personal agency. As a result, the TRIAD model serves not only as a psychological map but also as a guide to creating technologies that protect individuals' psychological integrity in an automated world.

\section{Discussion and Research Gaps}
\label{sec:discussion}

The research gaps outlined here target scholars, educators, practitioners, AI designers, researchers in human-computer interaction (HCI), and policy stakeholders who are interested in the long-term effects of widespread AI integration. Instead of capitalizing on short-term performance improvements, one must highlight the developmental, regulatory, and psychosocial paths that unfold over time. Our goal is to assist researchers in designing long-term studies, help educators and clinicians who use AI in their work, or treat patients who use AI, and support policymakers and technologists in creating protections that maintain cognitive independence, emotional growth, and social strength throughout life.

The preceding survey of the psychological triad illustrates that while AI offers significant scaffolding for human cognition and emotional support, it introduces risks of agency erosion and skill degradation. This section outlines the critical research gaps that must be addressed to ensure the sustainable, beneficial, and ethical integration of AI into human life.

\subsection{Cognitive and Behavioral Research Gaps}
Current research primarily focuses on short-term performance metrics, such as accuracy and speed, often neglecting the long-term cognitive costs of behavioral offloading \citep{Kos25, Bas25}. Key areas for future inquiry include:

    \subsubsection{Workforce Dynamics} Longitudinal studies are required to understand how the integration of AI into professional workflows impacts workforce dynamics and whether high-level problem-solving skills eventually erode through "routine-task" automation \citep{Bab24}. Further study is also needed for scenarios in which AI transitions from assistive tools to functional "peers", especially within a professional environment. Such work may investigate how role boundaries are enforced, authority structures, the evolution of accountability norms, and the organizational frameworks required to manage trust, oversight, and shared accountability in hybrid teams.
    
    \subsubsection{Therapeutic "Active Ingredients"} For clinical applications, research must determine whether success is driven by AI availability 24/7 or by specific personalization features. Stratified clinical trials are needed to evaluate the effectiveness in different severity levels of conditions, such as ASD or dementia subtypes.
    
    \subsubsection{Novelty Effects vs. Adherence} Future studies must move beyond initial "novelty effects" to assess long-term behavioral adherence and whether users maintain engagement with AI interventions over months or years. In addition to ongoing engagement, research should examine whether extended interaction with AI systems alters users' decision-making compared to automated guidance. As AI becomes more available and conversational, users might start assigning not just complex tasks but also small daily decisions, such as timing choices or planning routines. While this could reduce cognitive load, the long-term effects on autonomy, self-control, and decision-making confidence remain uncertain.
    
    \subsubsection{Cognitive development} How long-term AI exposure affects children's and adolescents' developing cognitive-emotional-behavioral systems is a significant unanswered question. According to the suggested triadic framework, AI output can systematically alter behavioral execution patterns (goal-directed vs. reactive), affect emotional assessments of competence and agency, and shift cognitive scope (e.g., toward higher-order analysis or surface-level completion). Repeated behavioral offloading may recalibrate attentional thresholds, persistence norms, and perceived locus of control during developmental stages when executive control, metacognition, and self-regulation are still developing. Therefore, longitudinal educational research is required to investigate whether AI-assisted learning reduces cognitive effort, dependency patterns, and diminished productive struggle, or whether it strengthens higher-order reasoning, knowledge transfer, and cognitive autonomy. Particular attention should be paid to sensitive developmental windows in which epistemic trust, motivational structures, and skill consolidation are still forming, as early recursive coupling between AI scaffolding and user regulation may have durable downstream effects on learning trajectories.
    
\subsection{Emotional and Ethical Research Gaps}
The emotional impact of AI is often measured in isolation, without robust human control groups, leading to a lack of clarity regarding the authenticity of AI-mediated support.

    \subsubsection{Comparative Authenticity} A critical gap is the absence of "venting-to-human" control groups in studies of AI emotional support. This would allow for a direct comparison of perceived social support and the long-term impact on human-to-human empathy.
    
    \subsubsection{Attachment and Development} It remains unclear how "frictionless" and sycophantic AI algorithms impact attachment models in early childhood. Research is needed to determine if these interactions disrupt healthy emotional development by removing the necessary "social friction" found in human relationships.
    
    \subsubsection{Positive Mental Health Indicators} Rather than focusing solely on symptom reduction (e.g., PHQ-9 or GAD-7 scores), future research should incorporate positive indicators such as overall life satisfaction and flourishing.
    
    \subsubsection{Artificial Intelligence Replacement Dysfunction (AIRD)} The future scale of job displacement resulting from AI deployment is unknown. However, whether unemployment due to AI use in professional settings is significantly lower than expected, occurs slowly over time, or is a larger-than-expected surge, mental health consequences warrant study as a related but distinct phenomenon. As uncertainty persists, so too will related psychological experiences. Comparative studies linking AI employment shifts to prior technology introductions may help define the issue's quality and scope. 
    
    \subsubsection{Psychosocial Development} Beyond individual emotions, a significant research gap involves how ongoing interactions with AI systems affect broader psychosocial development throughout life, especially in adolescence and young adulthood. Within the triadic framework, interactions with AI can shape how people evaluate emotions, such as perceived acceptance, validation, or social comparison. These interactions can change behavior patterns related to social engagement, such as replacing or adding to human relationships. They may also adjust how individuals think about trust, reciprocity, and personal control. During crucial developmental phases, when identity formation, independence negotiation, and building peer connections are essential, frequent exposure to smooth, one-sided, or excessively flattering AI systems could alter expectations for social feedback, conflict resolution, and effort in relationships. Long-term studies are needed to determine whether AI companionship boosts self-reflection and emotional expression, or whether it leads to social withdrawal, dependency, and reduced tolerance for uncertainty in social situations. Researchers should pay close attention to how AI systems affect the development of empathy, moral reasoning, trust in knowledge, and the formation of stable self-views among young people. The interaction between AI and the development of psychosocial ideas may have lasting effects on social skills and community participation.

    \subsubsection{Affective Interface Governance} A significant technological gap exists in the lack of design standards for how AI systems handle emotional tone, validation methods, and relationship framing. Current systems often focus on engagement and user satisfaction, but they lack clear protections against over-validation, dependency reinforcement, or manipulative emotional mirroring. Research is needed to figure out how emotional expressiveness, human-like features, and affirmation methods should be adjusted for different age groups and vulnerability levels. For example, easy empathy and constant validation might unintentionally change attachment expectations or lower tolerance for ambiguity in relationships. Future work should investigate ways to adjust emotional intensity, improve transparency in emotional simulation, and implement safeguards that protect user independence while reducing the risks of emotional overreliance or psychological manipulation.

    \subsubsection{Recursive Feedback Loops and Algorithmic Dependency Formation} A key technological gap involves the recursive feedback structures in adaptive AI systems. Many systems continually learn from user behavior, reinforcing interaction patterns that boost engagement or foster a sense of alignment. However, there is little research on how these reinforcement loops may increase emotional dependence, confirmation bias, or lower tolerance for disagreement over time. From an ethical perspective, we need studies to examine whether personalization engines unintentionally create emotional echo chambers, reinforce harmful coping patterns, or shift expectations around validation and agreement. Future work should examine how to introduce deliberate variability, friction, or balancing signals into adaptive systems to prevent unhealthy dependency while maintaining usability and support.


\subsection{Technology Focused Research Gaps}
Collectively, the future of human AI interactions will require active and orchestrated efforts on the technology front, as well, where research in AI systems' design should consider:

    \subsubsection{Human-Centered Interface Adaptation} A significant technological gap involves the limited integration of user characteristics into AI interface design. Current systems primarily focus on performance and engagement metrics, failing to account for demographic factors, age groups, digital skills, cognitive abilities, and areas of expertise. From a UI/UX perspective, future research should examine how adaptive interface designs can adjust levels of automation, transparency, explanation depth, and support based on user profiles. For instance, novice users or children may need more guidance, prompts, and cues to avoid relying on the system too soon. In contrast, expert users might benefit from more independence and flexible control options. Empirical studies are needed to find out how design choices affect attention, self-regulation, and long-term skill retention across different groups.

     \subsubsection{Mental Privacy and Bias} Ethically, there is a significant need to establish frameworks for "mental privacy" to protect biometric data that AI can use to infer internal emotional states without conscious disclosure. Additionally, probing for "data drift" and social biases within these systems remains a high priority.

    \subsubsection{Fostering Cognitive Autonomy} Future research should investigate the development of AI systems capable of calibrated intervention. Rather than providing exhaustive outputs by default, "scaffolded" systems could detect when a user is in a learning phase and intentionally limit assistance. Such a shift is essential to mitigate the risks of cognitive offloading and skill erosion, ensuring that users maintain the capacity for independent problem-solving in educational and professional environments.
    
   \subsubsection{Engineering Mindful Engagement} There is a critical need to explore the implementation of "productive friction" within digital interfaces. While current AI design prioritizes seamlessness, this often facilitates passive acceptance and diminishes critical thinking. Research should evaluate the impact of intentional "metacognitive pauses" requiring users to justify queries or verify generated evidence—to re-engage analytical processing and reduce the prevalence of reactive, unreflective interactions.
    
    \subsubsection{Prioritizing Affective and Social Well-being} A vital research direction involves the integration of emotional and behavioral sensing to detect maladaptive usage patterns. Systems could be designed to identify early indicators of over-attachment, emotional dependency, or social withdrawal. By recognizing these psychological shifts, AI could transition from a constant "digital crutch" to a supportive coach, actively encouraging real-world social reintegration and self-regulation when dependency thresholds are met.
    
    \subsubsection{Evaluating Longitudinal Human Impact} Finally, the field must move beyond cross-sectional observations to conduct long-term longitudinal studies. The cumulative impact of lifetime AI integration on human memory, social cognition, and agency remains largely unknown. Future investigations should focus on the "recovery of function," measuring the duration and effort required for individuals to regain lost skills after AI removal, to fully quantify the long-term psychological costs of prolonged reliance on algorithms.

    \subsubsection{Real-Time Cognitive State Estimation} 
    A significant technical barrier remains in developing a robust "sensor layer" capable of monitoring a user’s internal mental state by fusing variables such as typing rhythm, gaze patterns, and response latency. Successful implementation allows for "empathetic AI" that reduces user burnout and prevents cognitive overload by simplifying interfaces during high-stress periods. However, high-fidelity monitoring raises profound privacy concerns regarding "mental surveillance" and requires significant computational overhead to distinguish between task-related fatigue and external environmental stressors.

    \subsubsection{Dynamic Level of Automation (LoA) Controllers} 
    There is a lack of sophisticated control logic that can modulate the "degree of help" an AI provides based on the user’s evolving competence. This "adaptive scaffolding" ensures that AI acts as a catalyst for human growth rather than a replacement, effectively mitigating the skill erosion identified in pedagogical research. Conversely, poorly calibrated transitions between automation and manual control can lead to "automation shock," where a user is suddenly forced to take over a complex task they are no longer mentally prepared to handle, increasing the risk of critical errors.

    \subsubsection{Architectures for Productive Friction} 
    Current AI architectures are optimized for "frictionless" interaction, which often facilitates passive, unreflective user behavior. Engineering intentional "metacognitive speed bumps" can re-engage analytical processing, foster more critical evaluation of AI-generated content, and reduce the "autopilot" effect. The primary drawback is a potential degradation of user satisfaction; excessive or poorly timed friction may be perceived as a system malfunction or inefficiency, leading to user frustration and decreased adoption of the tool.

    \subsubsection{Longitudinal Competence Modeling} 
    We currently lack technical frameworks to track the psychological effects of AI integration over months or years through persistent "user-state models." Establishing these digital records enables "system health" audits that can proactively trigger skill-reinforcement sessions when a user's primary problem-solving abilities degrade beyond a safety threshold. However, the long-term storage of such sensitive psychological profiles creates a target for data breaches and raises ethical questions regarding who owns the "profile" of a user’s cognitive health and how it might be used by third parties.


\section{Conclusion}
\label{sec:conclusion}

The investigation of human--AI interactions through the lens of the psychological triad—cognition, behavior, and emotion—reveals a landscape defined by a profound duality. As AI systems transition from passive task-oriented tools to proactive, affective companions, they offer unprecedented opportunities for cognitive scaffolding, behavioral personalization, and emotional support. However, these benefits are inextricably linked to risks of cognitive offloading, the erosion of human agency, and the potential for emotional over-dependence. The findings of this review underscore that the impact of AI is rarely unidirectional; rather, it is mediated by the user’s pre-existing psychological state, the interaction context, and the design of the AI interface. To ensure that AI serves as a sustainable enrichment of human psychology rather than a replacement for it, the field must prioritize longitudinal research that examines the long-term effects of AI reliance on neuroplasticity and social-emotional development. 

Ultimately, this paper proposes that the goal of human-centric AI design should be a symbiotic relationship where technological augmentation preserves human autonomy. The TRIAD model introduced herein provides a necessary integrated framework to capture how cognitive processes, behavior patterns, and emotional assessments co-evolve within human--AI systems. By moving beyond isolated outcome measures toward a recursive, systems-level understanding, we can better address current research gaps in mental privacy, longitudinal skill retention, and cross-demographic validation. From an engineering and design perspective, this requires incorporating user traits, developmental stages, and interaction settings directly into the architecture of adaptive AI. By grounding technological advancements in these triadic psychological principles, we can move toward an era of AI integration that enhances the breadth and depth of the human experience while safeguarding fundamental psychological integrity and well-being.

\section*{CRediT authorship contribution statement}
All authors: Conceptualization, Methodology, Formal analysis, Writing - Original Draft, Writing - Review \& Editing.

\section*{Declaration of competing interest}
The authors declare no competing interests.

\section*{Declaration of generative AI and AI-assisted technologies in the manuscript preparation process}

During the preparation of this work, the authors used Google's Gemini to assist with LaTeX table formatting. After using this tool, the authors reviewed and edited the content as needed and take full responsibility for the content of the published article.

\end{document}